\documentclass[11pt,letterpaper]{article}
\usepackage{setspace}
\usepackage{color}
\usepackage{amsmath}
\usepackage{amsfonts}
\usepackage{subcaption}
\usepackage{comment}
\usepackage{booktabs}
\usepackage{verbatim}
\usepackage{amssymb}
\usepackage{graphicx}

\providecommand{\U}[1]{\protect\rule{.1in}{.1in}}
\begin{document}

\title{On the robustness of solitons crystals in the Skyrme model}
\author{Gonzalo Barriga$^{1,2}$, Fabrizio Canfora$^{3,1}$, Marcela Lagos$^{4}$%
, Mat\'ias Torres$^{1,2,5}$, Aldo Vera$^{4}$ \\
$^{1}$\textit{Centro de Estudios Cient\'{\i}ficos (CECS), Casilla 1469,
Valdivia, Chile,}\\
$^{2}$\textit{Departamento de F\'{\i}sica, Universidad de Concepci\'{o}n,
Casilla 160-C, Concepci\'{o}n, Chile}\\
$^{3}$\textit{Universidad San Sebastián, sede Valdivia, General Lagos 1163,}\\
\textit{Valdivia 5110693, Chile}\\
$^{4}$\textit{ Instituto de Ciencias F\'isicas y Matem\'aticas, Universidad
Austral de Chile,}\\ \textit{Casilla 567, Valdivia, Chile}\\
$^{5}$\textit{ Dipartimento di Fisica ``E. Pancini", Università di Napoli Federico II - INFN sezione di Napoli,}\\
\textit{Complesso Universitario di Monte S. Angelo Edificio 6, via Cintia, 80126 Napoli, Italy}\\
\\
{\small gobarriga@udec.cl, fabrizio.canfora@uss.cl, marcela.lagos@uach.cl,
matiatorres@udec.cl, aldo.vera@uach.cl}}
\maketitle

\begin{abstract}
In this work we analize how the inclusion of extra mesonic degrees of freedom
affect the finite density solitons crystals of the Skyrme model.
In particular, the first analytic examples of hadronic crystals at finite baryon density in
both the Skyrme $\omega $-mesons model as well as for the Skyrme $\rho $%
-mesons theory are constructed. These configurations have arbitrary
topological charge and describe crystals of baryonic tubes surrounded by a
cloud of vector-mesons. In the $\omega $-mesons case, it is possible to
reduce consistently the complete set of seven coupled non-linear field
equations to just two integrable differential equations; one ODE for the
Skyrmion profile and one PDE for the $\omega $-mesons field. This analytical
construction allows to show explicitly how the inclusion of $\omega $-mesons
in the Skyrme model reduces the repulsive interaction energy between
baryons. In the Skyrme $\rho $-mesons case, it is possible to construct
analytical solutions using a meron-type ansatz and fixing one of the
couplings of the $\rho $-mesons action in terms of the others. We show that,
quite remarkably, the values obtained for the coupling constants by
requiring the consistency of our ansatz are very close to the values used in
the literature to reduce nuclei binding energies of the Skyrme model without
vector-mesons. Moreover, our analytical results are in qualitative agreement
with the available results on the nuclear spaghetti phase.
\end{abstract}

\newpage

\tableofcontents


\section{Introduction}


\label{Intro} A detailed description of the phase diagram of Quantum
Chromodynamics (QCD henceforth) -especially at low temperature and finite
baryon density- is one of the greatest open challenges in theoretical
physics. It is usually assumed that it is not fruitful to analyze the
complex phase diagram of QCD in this regime with analytic tools (see \cite%
{newd3}, \cite{newd4}, \cite{newd5}, \cite{newd6}, and references therein)
and, consequently, smart numerical methods must be used. One of the most
remarkable features, which manifests itself when many baryons coexist within
a finite volume, is the appearance of the so-called \textit{nuclear pasta
phase} (see \cite{pasta1}, \cite{pasta2}, \cite{pasta2a}, \cite{pasta2b}, 
\cite{pasta3}, \cite{pasta4}, \cite{pasta5}, \cite{pasta6}, \cite{pasta7}, 
\cite{pasta8}, \cite{pasta9} and the nice up to date review \cite{pasta10}).
In such a phase ordered structures appear, in which most of the baryonic
charge is contained in regular shapes like thick baryonic layers (called 
\textit{nuclear lasagna}) or thick baryonic tubes (called \textit{nuclear
spaghetti}). Not surprisingly, due to the large number of strongly
interacting particles characterizing these nice regular structures, the
nuclear pasta phase is considered to be the prototype configuration where
analytic approaches are expected to fail.\footnote{%
For numerical results on multi-solitons crystals see \cite{aprox0}, \cite%
{aprox1}, \cite{aprox2}, \cite{aprox3}, \cite{aprox4}, \cite{aprox5}, \cite%
{aprox6}, \cite{aprox7}, \cite{aprox8}, \cite{aprox9}, \cite{aprox10} and
references therein.} Even more, in such phases the numerical analysis are
quite challenging and, as the above references show clearly, very high
computing power is required.

In this paper we will analyze the appearance of these structures at finite
density as analytical solutions of the Skyrme theory (in particular when
this model is coupled to vector-mesons), which at leading order in the t'
Hooft expansion \cite{Gerard} represents the low energy limit of QCD.

The Skyrme theory \cite{skyrme} is described by a bosonic action for a $%
SU(N) $-valued scalar field $U$ (being the two-flavors case $U(x)\in SU(2)$
the most frequently studied), whose small excitations represent pions while
the topological solitons of the theory are interpreted as baryons \cite%
{Lizzi}, \cite{shifman1}, \cite{shifman2}, \cite{witten0}, \cite{ANW}. It is
worth to emphasize that the interest on the Skyrme model goes far beyond QCD
since it has been applied in astrophysics \cite{ref1}, Bose-Einstein
condensates \cite{ref2}, nematic liquids \cite{ref3}, magnetic structures 
\cite{ref4} and condensed matter physics \cite{ref5} (see also \cite%
{BalaBook} and \cite{manton}). However, as it is well known since the
eighties, the predictions of the Skyrme theory on the neutron-proton mass
difference, the baryon resonances, the nuclei binding energies, the
electromagnetic form factors and the matrix element of the singlet axial
current, are not in excellent agreement with experiments. Fortunately, there
is a very natural way to improve the theoretical results mentioned above,
and it is including more families of vector-mesons such as the $\omega $%
-mesons and the $\rho $-mesons to the Skyrme action \cite{[8]p}, \cite{[9]p}%
, \cite{[10]p}, \cite{[11]p}, \cite{[12]p}, \cite{sakurai}, \cite{Machleidt}
(see also \cite{m1}, \cite{m2} and \cite{m3}). The first part of this work
it is devoted to the construction of analytic solutions of the Skyrme $%
\omega $-mesons theory and the second part to the Skyrme $\rho $-mesons
theory.

In Skyrme-like models containing vector-mesons the field equations appears
to be tremendously more complicated than the ones of the Skyrme model alone,
but (besides the already mentioned results in \cite{[8]p}, \cite{[9]p}, \cite%
{[10]p}, \cite{[11]p}, \cite{[12]p}, \cite{sakurai}, \cite{Machleidt}) there
are further sound reasons to include the $\omega $-mesons and $\rho $-mesons
in the game nevertheless. First, in \cite{AN}, the authors pointed
out that it is possible to stabilize the Skyrmions without the need to
consider the Skyrme term by coupling the baryonic current to the $\omega $%
-mesons. Secondly, the ``stabilizing role" of the $\omega $%
-mesons becomes more and more important as the baryonic charge increases
(see \cite{BPS}, \cite{BPS1}, \cite{BPS2}, \cite{BPS3}, \cite{BPS4} and
references therein). Consequently, as we are interested in describing
configurations with high topological charge (which is a necessary condition
in the formation of nuclear pasta), it is extremely important to include the
effects of the $\omega $-mesons in our analysis. On the other hand, the
Skyrme model produces nuclei binding energies that are larger than the
experimental ones, and the clustering structure of light nuclei in the
Skyrme model is also not optimal. However, both of these problems can be
solved by including the next lightest Isospin $1$ meson, namely, the $\rho $%
-mesons \cite{NS1}, \cite{NS2}.

The main goal of the present paper is to construct analytic configurations
of the Skyrme vector-mesons theory representing crystals of baryons
surrounded by a clouds of $\omega $-mesons and $\rho $-mesons,
suitable for describing the nuclear spaghetti phase. The physical
motivation is to reach a deeper understanding with some analytic control on
how the vector-mesons affect the complex configurations of nuclear pasta
appearing at finite baryon density. An important byproduct of our analysis
will be to disclose the importance of the inclusion of the vector-mesons in
the nuclear pasta phase.\footnote{Although the half-Skyrmion configuration \cite{half}
gives a good numerical description at high density of baryonic matter,
it does not describe phases in which the nucleons loose their individuality,
i.e, phases like nuclear pasta, where nucleons are ``melted" at finite volume forming ordered patterns.}

In order to achieve these goals, we will generalize the methods introduced
in \cite{56}, \cite{56b}, \cite{56a0}, \cite{56a}, \cite{56a1}, \cite{56a2}, 
\cite{56b1}, \cite{56c}, \cite{crystal1}, \cite{crystal2}, \cite{crystal3}, 
\cite{crystal3.1}, \cite{crystal4}, \cite{crystal4.1} to the Skyrme vector-mesons theory
(see also \cite{firstube}, \cite{firsttube2}, \cite{gaugsksu(n)}, \cite{Federica}, \cite{Daniel} and \cite{Canfora:2021lbe}).
In fact, the above references have allowed the construction of several analytic
and topologically non-trivial solutions of the Skyrme model. From the
viewpoint of the goals of the present work, the most relevant configurations
analyzed in those references corresponds to ordered baryonic arrays in which
(most of) the topological charge and total energy are concentrated within
tube-shaped regions\footnote{%
In \cite{Jackson} and \cite{nittain}, numerical string shaped solutions in
the Skyrme model with mass term have been constructed. However, those
configurations have a zero topological density (and they are expected to
decay into pions). The configurations analyzed in the present paper are
topologically non-trivial and therefore can not decay into those of \cite%
{Jackson} and \cite{nittain}.}. In particular, the similarity of the contour
plots in \cite{crystal1} with the spaghetti-like configurations found
numerically (see the plots in \cite{pasta1}, \cite{pasta2}, \cite{pasta2a}, 
\cite{pasta2b}, \cite{pasta3} and \cite{pasta10}) are extremely encouraging,
and strongly support the viability of the present analytic approach\footnote{%
Using this framework, in \cite{gaugsksu(n)} it has been possible to compute
the shear modulus of lasagna configurations obtaining good agreement with 
\cite{pasta5} and \cite{pasta9}. Moreover, in \cite{crystal4.1}, the very
interesting characteristics of the electromagnetic field generated by
nuclear spaghetti have been analyzed explicitly.}. The analytic results
described in the following sections allow to compute several relevant
physical quantities such as the distribution of the $\omega $-mesons around
the peaks of the baryon density and the ``shielding" effect
of the $\omega $-mesons in the repulsive Skyrmion-Skyrmion interactions.

The case of the $\rho$-meson Skyrme theory of \cite{NS1}, \cite{NS2} is the
most complex due to the very intricate non-linear interactions between the
Skyrmions and the $\rho $-mesons. Nevertheless, in \cite{NS1} and \cite{NS2},
the authors showed numerically that there are (at least two) possible
choices of the many coupling constants of the theory giving rise to nuclei
binding energy in very good agreement with experiments. In order to deal
with the fifteen non-linear coupled field equations of the $\rho $-meson
Skyrme theory we use the ansatz of \cite{crystal1} for the $U$ field and a meron-like
ansatz for the $\rho $-mesons (which works extremely well in the Yang-Mills
case: see \cite{Alfaro}, \cite{meron0}, \cite{meron1}, \cite{meron2}, \cite%
{meron3} and \cite{meron4}). Of course, one may wonder whether the extremely
complicated non-linear field equations of the $\rho $-meson Skyrme theory of 
\cite{NS1} and \cite{NS2} can be dealt with a meron-like ansatz, given the fact
that there is no non-Abelian gauge symmetry in the $\rho $-mesons case.

In fact, the present approach is surprisingly effective in this case as
well. One can fix one of the coupling constants of \cite{NS1} and \cite{NS2} in
terms of the other coupling constants of the theory by requiring the
consistency of our ansatz (namely by requiring the solvability of the field
equations in topologically non-trivial sectors of high baryonic charge). At
a first glance, one could think that such a requirement is a bit artificial.
In particular, there is no obvious reason why the values of the coupling
constants arising insisting that the present ansatz must work in the $\rho $%
-mesons Skyrme theory should in any way be related with the values of the
coupling constants of \cite{NS1} and \cite{NS2} (chosen by the authors in order
to achieve agreement with the nuclei binding energies). Nevertheless, quite
remarkably, the choice of the coupling constants which makes the present
framework suitable to deal with the $\rho $-mesons Skyrme theory is very
close to the results in \cite{NS1} and \cite{NS2}. With the inclusions of
the vector mesons, our analytical results are in qualitative agreement with
the available results on the nuclear spaghetti phase.

The paper is organized as follows. In Section 2 we construct analytical
solutions for the Skyrme $\omega$-mesons theory, and we show that these
configurations can be interpreted as a lattice of baryonic tubes surrounded
by $\omega$-mesons. Also we compute the effect of the $\omega$-mesons on the
repulsive interaction energy. In Section 3, using a similar approach, we
present analytical solutions of the Skyrme $\rho$-mesons theory showing that
the inclusion of the $\rho$-mesons ``shield" the interaction between the
baryons on the crystal. Section 4 is devoted to the conclusions.

In our convention $c=\hbar =1$, Greek indices run over the space-time with
mostly plus signature and Latin indices are reserved for those of the
internal space.


\section{Crystals of Baryons with a cloud of $\omega$-mesons}


\label{section2}

In this section we will construct analytical solutions describing crystals
of baryonic tubes in the Skyrme $\omega$-mesons theory.

\subsection{The Skyrme $\protect\omega$-mesons theory}

The action for the $SU(2)$-Skyrme model coupled with the $\omega $-mesons 
\cite{AN} is given by 
\begin{align}
I(U,\omega ) \ = \ \int \sqrt{-g} d^{4}x  \biggl[ & \frac{K}{4}\text{Tr}\left( R_{\mu
}R^{\mu }+\frac{\lambda }{8}F_{\mu \nu }F^{\mu \nu }\right) -\frac{M_{\pi
}^{2}}{2}\text{Tr}\left( U+U^{-1}\right)  \notag \\
& -\frac{1}{4}S_{\mu \nu }S^{\mu \nu} -\frac{1}{2}M_{\omega }^{2}\omega
_{\mu }\omega ^{\mu }+g_{0}J_{\mu }\omega ^{\mu }\biggl] \ ,  \label{I1}
\end{align}
\begin{equation*}
R_{\mu }=U^{-1}\nabla _{\mu }U=R_{\mu }^{a}t_{a}\ ,\quad F_{\mu \nu
}=[R_{\mu },R_{\nu }]\ ,\quad S_{\mu \nu }=\partial _{\mu }\omega _{\nu
}-\partial _{\nu }\omega _{\mu }\  ,
\end{equation*}%
where $U(x)\in SU(2)$, $\omega _{\mu }$ is a $4$-vector, $g$ is the metric
determinant, $\nabla _{\mu }$ is the Levi-Civita covariant derivate and $%
t_{a}=i\sigma _{a}$ are the generators of the $SU(2)$ Lie group, being $%
\sigma _{a}$ the Pauli matrices. The Skyrme couplings $K$ and $\lambda $ as
well as the $\omega$-mesons coupling $g_{0}$ are positive constant fixed
experimentally, while $M_{\pi }$ and $M_{\omega }$ correspond to the pions
and $\omega $-mesons mass, respectively.
Here the parameters $K$ and $\lambda$ are related to the meson decay coupling constant $F_{\pi}$ and the Skyrme coupling $e$
via $F_{\pi}=2 \sqrt{K}$ and $K \lambda e^{2}=1$, where $F_{\pi}=141 \text{MeV}$ and $e=5.45$. Thus, the energy, $\int d^3 x \sqrt{-g} T^{00}$,
is in units of $F_\pi/e$ \text{MeV} (all the energy plots of the present paper will be given in terms of these units).

The pions and $\omega$-mesons interact in Eq. \eqref{I1} through a term
proportional to the topological current, $J_{\mu}$, which is defined as 
\begin{equation}  \label{Bcurrent}
J^{\mu}=\epsilon^{\mu\nu\alpha\beta}\text{Tr}(R_{\nu}R_{\alpha}R_{\beta}) \ .
\end{equation}
Integrating the temporal component of the topological current defined above
on a space-like hypersurface one obtain the topological charge 
\begin{gather}  \label{Bcharge}
B=\frac{1}{24\pi^{2}}\int_{\Sigma}J^{0}\ , \qquad J^{0}=\epsilon^{ijk}\text{%
Tr}\left[R_{i}R_{j}R_{j} \right] \ ,
\end{gather}
which in the Skyrme model is identified as the baryonic number. 

At this point is it important to emphasize the relations between the Skyrme $\omega$-mesons theory and the Walecka model \cite{walecka2}.
In the study of compact stars, the Walecka model is a very useful theory of nucleons and two mesons: the scalar meson $\sigma$ and the $\omega$-vector meson
(see \cite{walecka2}, \cite{walecka3} and \cite{walecka1}).
The Lagrangian density of the Walecka model is given by,
\begin{align}
\mathcal{L}_{W}=\bar{\psi}\left[i \gamma_{\mu}\left(\partial^{\mu}+i g_{\omega} \omega^{\mu}\right)-\left(m_{0}-g_{\sigma} \sigma\right)\right] \psi+\frac{1}{2}\left(\partial_{\mu} \sigma \partial^{\mu} \sigma-m_{\sigma}^{2} \sigma^{2}\right)-\frac{1}{4} \omega_{\mu \nu} \omega^{\mu \nu}+\frac{1}{2} m_{\omega}^{2} \omega_{\mu} \omega^{\mu} \ . 
\end{align}
Here the $\omega$-vector meson is coupled to the nucleons via a minimal coupling, while in the case of the Skyrme model is via the topological current. 
The present inhomogeneous topologically non-trivial configurations manifest clear similarities with the ones appearing in the nonhomogeneous phase of the Walecka model
(see Chapter 5 in \cite{walecka3}). Indeed, despite the fact that our analytic solutions live in three spatial dimensions, one of the profiles, namely $\alpha$,
can be expressed in terms of inverse elliptic functions, similar to what happens with the solutions in the Walecka model \cite{walecka3}.
Thus, in a sense, the presence of the Skyme field supports the inhomogeneous phase of the Walecka model (see \cite{walecka4}, \cite{walecka5}, \cite{walecka6} and references therein). 

In the next section we will see that the boundary conditions for the soliton profile emerge naturally
by requiring that the topological charge be an integer. 

The variation of the action in Eq. \eqref{I1} w.r.t the fundamental fields $%
U $ and $\omega_{\mu}$ leads to the following field equations 
\begin{align}   \label{EqSkyrme}
\nabla_{\mu}\left(R^{\mu}+\frac{\lambda}{4}[R_{\nu},F^{\mu\nu}]\right)+\frac{%
M_{\pi}^2}{K}(U-U^{-1})+6\frac{g_{0}}{K}\epsilon^{\alpha\nu\lambda\rho}
\nabla_{\nu}(\omega_{\alpha})R_{\lambda}R_{\rho}&=0 \ , \\
\nabla_{\nu}S^{\nu\mu}-M_{\omega}^{2}\omega^{\mu}+g_{0}J^{\mu}&=0 \ . \label{Eqomega}
\end{align}
Eqs. \eqref{EqSkyrme} and \eqref{Eqomega} are, in general, a set of seven
coupled non-linear partial differential equations. One of the main results
of the present work is that, despite the complexity of the above system,
this can be solved analytically using an appropiate ansatz, as we will see
below.

The energy-momentum tensor of the theory corresponds to 
\begin{align}
T_{\mu \nu} =& -\frac{K}{2} \text{Tr}\left[R_{\mu} R_{\nu}-\frac{1}{2}
g_{\mu \nu} R^{\alpha} R_{\alpha}+\frac{\lambda}{4}\left(g^{\alpha \beta}
F_{\mu \alpha} F_{\nu \beta}-\frac{1}{4}g_{\mu \nu} F_{\sigma \rho}
F^{\sigma \rho}\right)+\frac{M_\pi^2}{K}g_{\mu \nu}(U+U^{-1})\right]  \notag
\\
& +S_{\mu \alpha} S_{\nu}^{\alpha}-\frac{1}{4} S_{\alpha \beta} S^{\alpha
\beta} g_{\mu \nu}+M_{\omega}^2\left(\omega_{\mu}\omega_{\nu}-\frac{1}{2}%
g_{\mu\nu}\omega_{\alpha}\omega^{\alpha}\right)-g_{0}\left(J_{\mu}\omega_{%
\nu}+J_{\nu}\omega_{\mu}-g_{\mu\nu}J_{\alpha}\omega^{\alpha}\right) \ .
\label{Tmunu1}
\end{align}

\subsection{The Ansatz}

We will construct analytical solutions that describe states of
multi-solitons at finite density, so we consider as a starting point the
metric of a box whose line element\footnote{Here the coordinates
$\{r,\theta,\phi\}$ represent Cartesian coordinates and they must not be confused with spherical coordinates.} is 
\begin{equation}
ds^{2}=-dt^{2}+L^{2}(dr^{2}+d\theta ^{2}+d\phi ^{2})\ ,  \label{Box}
\end{equation}%
where $L$ is a constant representing the lenght of the box where the
solitons are confined. The adimensional coordinates $\{r,\theta ,\phi \}$
have the following ranges 
\begin{equation}
0\leq r\leq 2\pi \ ,\quad 0\leq \theta \leq \pi \ ,\quad 0\leq \phi \leq
2\pi \ ,  \label{ranges}
\end{equation}%
so that the volume available for the solitons is $V=4\pi ^{3}L^{3}$.

We parameterize the Skyrme field $U(x)$ as usual for an element of the $%
SU(2) $ group, namely 
\begin{equation}  \label{U1}
U^{\pm1}(x^\mu)= \cos(\alpha) \mathbf{1}_{2\times2} \pm \sin(\alpha)
n^{a}t_{a}\ ,
\end{equation}
\begin{equation}  \label{U2}
n^{1}=\sin\Theta\cos\Phi\ ,\quad n^{2}=\sin\Theta\sin\Phi\ ,\quad
n^{3}=\cos\Theta\ \ ,
\end{equation}
where $\mathbf{1}_{2\times2}$ is the $2\times2$ identity matrix and $%
\alpha=\alpha(x^\mu)$, $\Theta=\Theta(x^\mu)$ and $\Phi=\Phi(x^\mu)$ are the
three degrees of freedom of the $U(x)$ field. With the parameterization in
Eqs. \eqref{U1} and \eqref{U2} the topological charge density in Eq. %
\eqref{Bcharge} becomes 
\begin{equation}  \label{B01}
J^{0} = -12\left(\sin ^{2} \alpha \sin \Theta\right) d \alpha \wedge d
\Theta \wedge d \Phi\ .
\end{equation}
From the above expression it follows that, in order to have non-trivial
topological configurations, we must demand that $d \alpha \wedge d \Theta
\wedge d \Phi \neq 0$. This implies the necessary (but not sufficient)
condition that $\alpha$, $\Theta$ and $\Phi$ must be three indepedent
functions. The existence of arbitrary topological charge of our solutions
will be revealed later when appropriate boundary conditions be imposed.

The strategy introduced in \cite{crystal1} and \cite{crystal2} for the
Skyrme-Maxwell case provides with a very efficient ansatz which reduces the
complete set of seven coupled non-linear field equations to just two
integrable equations (one ODE for the Skyrmion profile and one PDE for the
Maxwell potential) keeping alive the topological charge. Remarkably, such a
strategy can be extended to the Skyrme $\omega$-mesons theory as follows.
Firstly, the functions $\alpha $, $\Theta $ and $\Phi $ must be choosen as 
\begin{equation}  \label{Uexplicit}
\alpha =\alpha (r)\ ,\quad \Theta =q\theta \ ,\quad \Phi =p\left( \frac{t}{L}%
-\phi \right) \ ,\quad q=2v+1\ ,\ \ v\in \mathbb{Z}\ .
\end{equation}%
The above ansatz is a very convenient choice for (at least) two reasons. The
first one is a practical reason, since it is straightforward to verify that
Eq. \eqref{Uexplicit} satisfies the following identities 
\begin{equation*}
\nabla _{\mu }\Phi \nabla ^{\mu }\alpha =\nabla _{\mu }\alpha \nabla ^{\mu
}\Theta =\nabla _{\mu }\Phi \nabla ^{\mu }\Phi =\nabla _{\mu }\Theta \nabla
^{\mu }\Phi =0\ ,
\end{equation*}%
which greatly simplify the field equations in Eq. \eqref{EqSkyrme}. The
second reason is because Eq. \eqref{Uexplicit} allows to avoid the Derrick's
theorem \cite{Derrick} due to the time dependence of the $U$ field. It is
also possible to verify that although the $U$ field depends explicitly on
time the energy density is static, in such a way that the solutions
constructed here are in fact topological solitons with finite energy.

The construction in \cite{crystal1} and \cite{crystal2} also suggest the
following ansatz for the $\omega$-mesons: 
\begin{equation}  \label{omega}
\omega _{\mu }=(u,0,0,-Lu)\ ,\qquad u=u(r,\theta )\ .
\end{equation}
The above expression in Eq. \eqref{omega} is also very convenient because it
satisfies the following relations 
\begin{equation*}
\epsilon ^{\alpha \nu \lambda \rho }\nabla _{\nu }\left( \omega _{\alpha
}\right) \text{Tr}[R_{\lambda }R_{\rho }]=0 \ , \qquad \omega _{\mu }\omega
^{\mu }=0 \ ,
\end{equation*}%
which not only significantly reduce the $\omega$-mesons field equations in
Eq. \eqref{Eqomega}, but also allows to decouple the contribution of the $%
\omega$-mesons from the Skyrme field equations in Eq. \eqref{EqSkyrme}.

\subsection{Analytical solutions}

Replacing the ansatz introduced in Eqs. \eqref{Box}, \eqref{U1}, \eqref{U2}, %
\eqref{Uexplicit} and \eqref{omega} into Eq. \eqref{EqSkyrme}, the set of
three non-linear differential Skyrme equations are reduced to only one first
order ODE for the profile $\alpha $, namely 
\begin{equation}
\partial _{r}\left[ Y(\alpha )\frac{\left( \alpha^{\prime} \right) ^{2}}{%
2}-V(\alpha )-E_{0}\right] =0\ ,  \label{Eqalpha}
\end{equation}%
being $\alpha^{\prime}=\frac{\partial \alpha}{\partial r}$, $E_{0}$ an integration constant (fixed by the boundary conditions, as
we will see below) and where we have defined the functions $Y(\alpha )$ and $%
V(\alpha )$ as follows 
\begin{equation}
Y(\alpha )=2(q^{2}\lambda \sin ^{2}(\alpha )+L^{2})\ ,\qquad V(\alpha )=-%
\frac{1}{2}L^{2}q^{2}\cos (2\alpha )+\frac{4L^{4}M_{\pi }^{2}\cos (\alpha )}{%
K}\ .  \label{YandV}
\end{equation}%
The equation for the profile $\alpha $, that can be conveniently written as 
\begin{equation}
\frac{d\alpha }{\eta (\alpha ,E_{0})}=\pm dr\ ,\quad \eta \left( \alpha
,E_{0}\right) =\frac{\left[ 2\left( E_{0}+V(\alpha )\right) \right] ^{1/2}}{%
Y(\alpha )^{1/2}}\ ,  \label{alpha}
\end{equation}%
can be solved analytically in terms of Elliptic Functions. Note that Eq. %
\eqref{Eqalpha} does not depend on the potential $u(r,\theta )$, i.e, it is
completely decoupled from the $\omega $-mesons due to the good properties of
the ansatz described above. In fact, the ansatz in Eqs. \eqref{U1}, %
\eqref{U2} and \eqref{Uexplicit} (that can be called \textquotedblleft 
\textit{generalized hedgehog ansatz}") has advantages with respect to the
original spherical hedgehog ansatz introduced by Skyrme in \cite{skyrme}.
Firstly, it allows to reduce the complete set of Skyrme equations to a first
order equation which can be solved analytically instead of numerically.
Secondly, it leads to configurations with arbitrary topological charge and,
moreover, is not restricted to spherical symmetry.

Similarly, the four field equations for the $\omega $-mesons in Eq. %
\eqref{Eqomega} are also reduced to just one equation but, in this case, a
partial differential equation for the function $u(r,\theta )$ introduced in
Eq. \eqref{omega}. Then, the $\omega $-mesons profile satisfies a
two-dimensional Poisson equation with a source term provided by the Skyrmion
profile, that is 
\begin{equation}
\left( -\frac{\partial ^{2}}{\partial r^{2}}-\frac{\partial ^{2}}{\partial
\theta ^{2}}+M_{\omega }^{2}L^{2}\right) u=J_{\text{eff}}\ ,\qquad J_{\text{%
eff}}=12g_{0}pqL^2 \sin (q\theta )\sin ^{2}(\alpha )\alpha ^{\prime }\ .
\label{Equ}
\end{equation}%
At this point it is important to highlight a crucial difference between this
reduction and what happens in the case where the non-linear sigma model
(NLSM) or the Skyrme model are coupled to the Maxwel theory. While in the
Skyrme-Maxwell (or NLSM-Maxwell) case the equations for the electromagnetic
field are reduced to a two-dimensional periodic Schr\"{o}dinger equation
that requires a numerical treatment (see \cite{crystal2} and \cite{crystal3}%
), in the present case the coupling with the $\omega $-mesons is simpler
since Eq. \eqref{Equ} can also be directly solved. In fact, the solution
of Eq. \eqref{Equ} can be obtained as
\begin{equation}
u(\vec{x})=\int d\vec{x}^{\prime }G(\vec{x},\vec{x}^{\prime })J_{\text{eff}}(%
\vec{x}^{\prime })\ ,\qquad (-\nabla _{\vec{x}}^{2}+M_{\omega }^{2}L^{2})G(%
\vec{x},\vec{x}^{\prime })=\delta (\vec{x}-\vec{x}^{\prime })\ ,
\label{solu}
\end{equation}%
where $x=(r,\theta )$.

Summarizing, using the generalized hedgehog ansatz for the Skyrme field in
addition to a null-vector as ansatz for the $\omega$-field, we are able to
reduce the total set of seven non-linear coupled differencial field
equations to only two integrable equations.

\subsection{Topological charge and energy density}

Plugging the ansatz in Eqs. \eqref{U1}, \eqref{U2} and \eqref{Uexplicit}
into Eq. \eqref{Bcharge}, the topological charge density of the matter field
reads 
\begin{equation*}
J^{0}=3pq\sin (q\theta )\frac{\partial }{\partial r}(\sin (2\alpha )-2\alpha
)\ .
\end{equation*}%
Since the baryon number must be an integer, it is straightforward to verify that for this purpose 
one needs to impose the following boundary condition on the soliton profile $\alpha (r)$: 
\begin{equation}
\alpha (2\pi )-\alpha (0)=n\pi \ ,  \label{bc}
\end{equation}%
with $n$ an integer number. Therefore, integrating in the ranges defined in
Eq. \eqref{ranges}, the topological charge takes the value 
\begin{equation}
B=-np\ ,
\end{equation}%
where we have used the fact that $q$ is an odd number, as specified in the
ansatz in Eq. \eqref{Uexplicit}. From Eqs. \eqref{ranges}, \eqref{alpha} and %
\eqref{bc} it follows that the integration constant $E_{0}$ is determined by the relation 
\begin{equation}
n\int_{0}^{\pi }\frac{d\alpha }{\eta \left( \alpha ,E_{0}\right) }=2\pi \ .
\label{E0n}
\end{equation}%
Equation \eqref{E0n} is an equation for $E_{0}$ in terms of $n$ that always has a
real solution when $E_{0}+V(\alpha )>0$, which implies that $E_{0}$ is
bounded from below 
\begin{equation}
E_{0}>\frac{L^{2}q^{2}}{2}+4\frac{L^{4}M_{\pi }^{2}}{K}\ ,  \label{ncfs}
\end{equation}%
so that, for given values of $q$ and $L$, the integration constant $E_{0}$
determines the value of the $\alpha $ profile for the boundary conditions
defined in Eq. \eqref{bc}.

\subsubsection{A necessary condition for stability}

When the field equations reduce to a single equation for the profile in a
topologically non-trivial sector (as in the present case) one says that
``\textit{the hedgehog property holds}". In many situations
(although not always, see for detailed discussions \cite{shifman1}, \cite%
{shifman2} and references therein) the perturbations which could more easily
lead to a decrease in the energy of the system are those perturbations of
the profile which keep the hedgehog property. In the present case, these
dangerous perturbations are of the following form:%
\begin{equation}
\alpha \rightarrow \alpha +\varepsilon \xi \left( r\right) \ ,\ \ \
0<\varepsilon \ll 1\ ,  \label{pert}
\end{equation}%
where $\alpha $ is the Skyrmion profile of the background solution, which do
not change the $SU(2)$ Isospin degrees of freedom. It is easy to see that
the linearized Skyrme field equations for $\xi \left( r\right) $ in Eq. (\ref%
{pert}) always has the following zero-mode: $\xi \left( r\right) =\partial
_{r}\alpha $. From here one can deduce the constraint in Eq. (\ref{ncfs}).
If the above condition is satisfied the zero mode $\xi \left( r\right)
=\partial _{r}\alpha $ has no node, hence the solution is stable under these
perturbations.

\subsubsection{Interacting Baryons surrounded by $\omega$-mesons}

Evaluating the ansatz defined in Eqs. \eqref{Box}, \eqref{U1}, \eqref{U2}, %
\eqref{Uexplicit} and \eqref{omega} in Eq. \eqref{Tmunu1}, the energy
density ($\mathcal{E}=T_{00}$) of the configurations constructed above turns
out to be
\begin{equation}
\mathcal{E} \ = \ \mathcal{E}_{\text{Sk}}+\mathcal{E}_{\omega }+\mathcal{E}_{%
\text{int}}\ ,
\end{equation}%
where we have defined 
\begin{align}
\mathcal{E}_{\text{Sk}}\ =\ & \frac{K\alpha ^{\prime 2}\left(L^{2}+\lambda
\sin ^{2}(\alpha )\left[q^{2}+ 2p^{2}\sin ^{2}(q\theta )\right] \right) }{%
2L^{4}}  \notag \\
& +\frac{K\sin ^{2}(\alpha )\left( L^{2}\left[q^2+2 p^2\sin ^{2}(q\theta)%
\right] +2\lambda p^{2}q^{2}\sin^2(q\theta)\sin^2\alpha\right) }{2L^{4}}%
+2M_{\pi }^{2}\cos (\alpha )\ , \\
\mathcal{E}_{\omega }\ =\ & \frac{(\partial _{\theta }u)^{2}+(\partial
_{r}u)^{2}}{L^{2}}+M_{\omega }^{2}u^{2}\ , \\
\mathcal{E}_{\text{int}}\ =\ & -24g_{0}pq\sin (q\theta )\sin^2(\alpha)\alpha
^{\prime }u\ .
\end{align}%
Note that the energy density (in general, the energy-momentum tensor) and the the topological charge density do not depend
on the cooordinate $\phi$ while they do depend on $r$ and $\theta$; that is why
these configurations describe the nuclear spaghetti phase.

\begin{figure}[!ht]
\centering
\includegraphics[scale=.42]{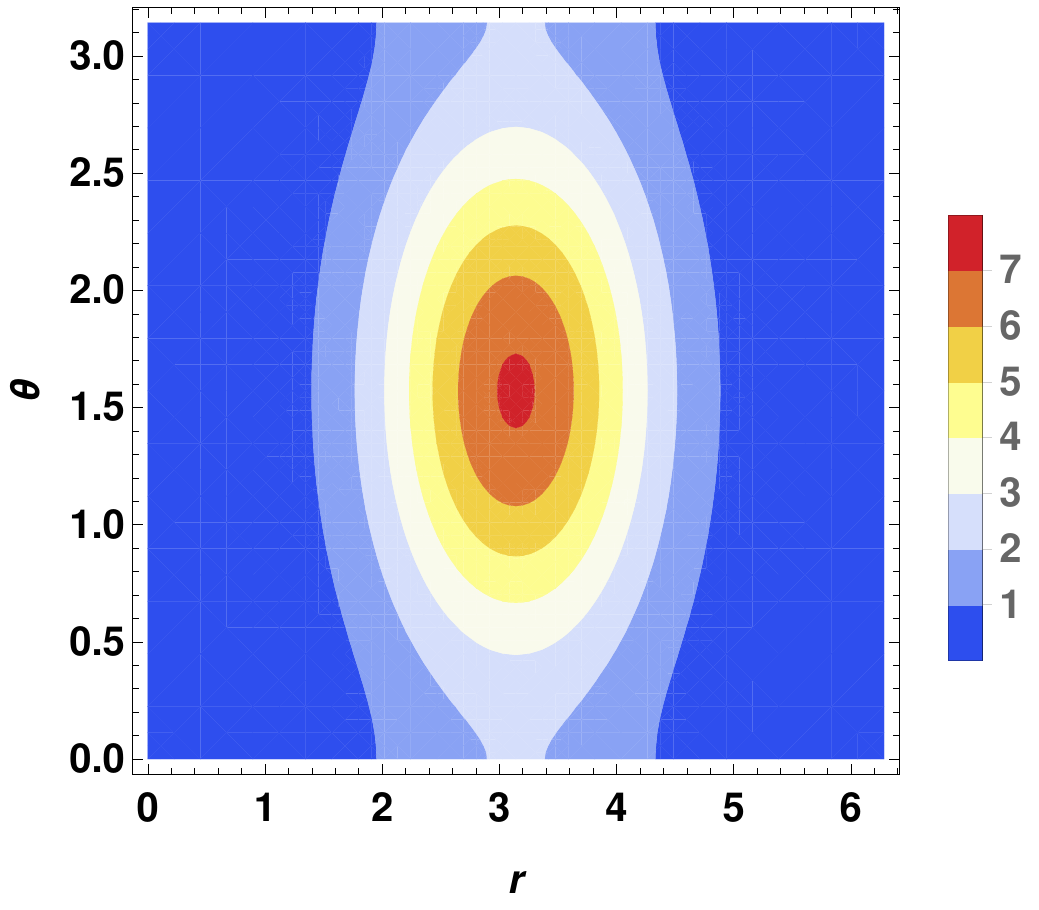}\qquad%
\includegraphics[scale=.42]{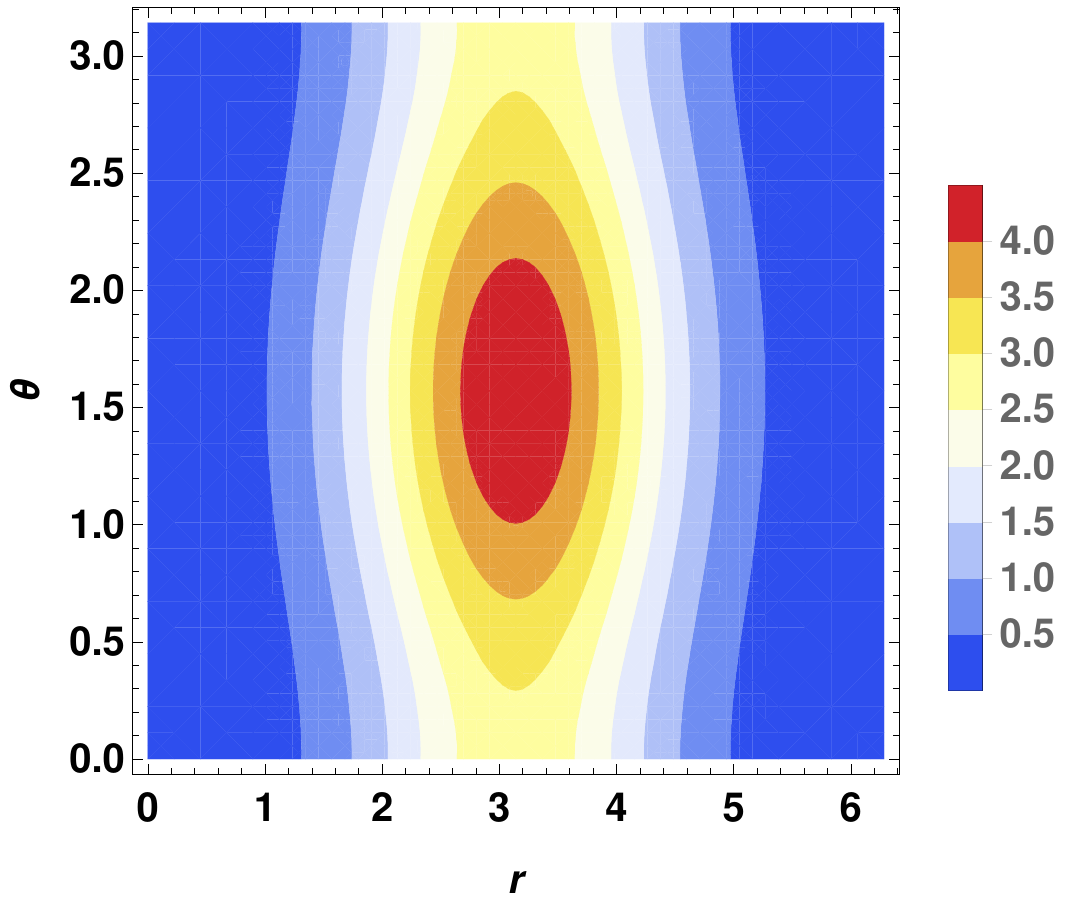} \qquad \qquad \qquad \qquad \qquad %
\includegraphics[scale=.42]{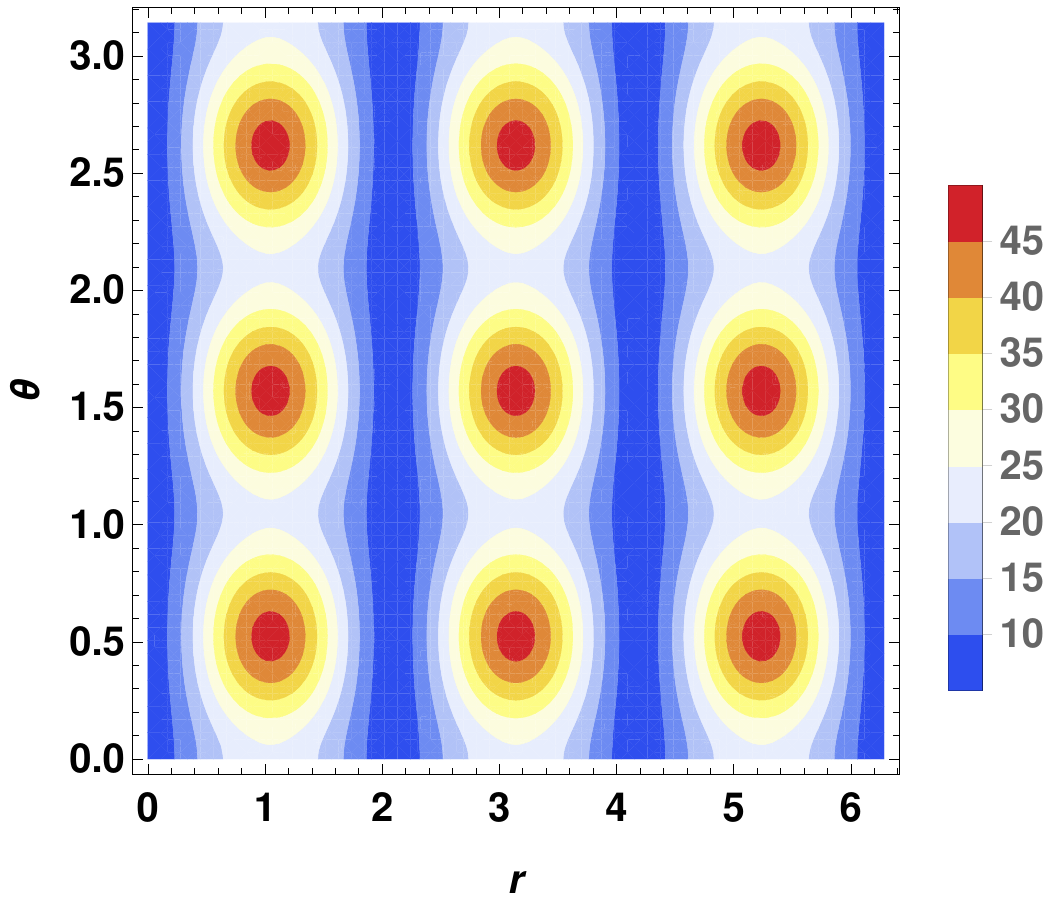}\qquad%
\includegraphics[scale=.42]{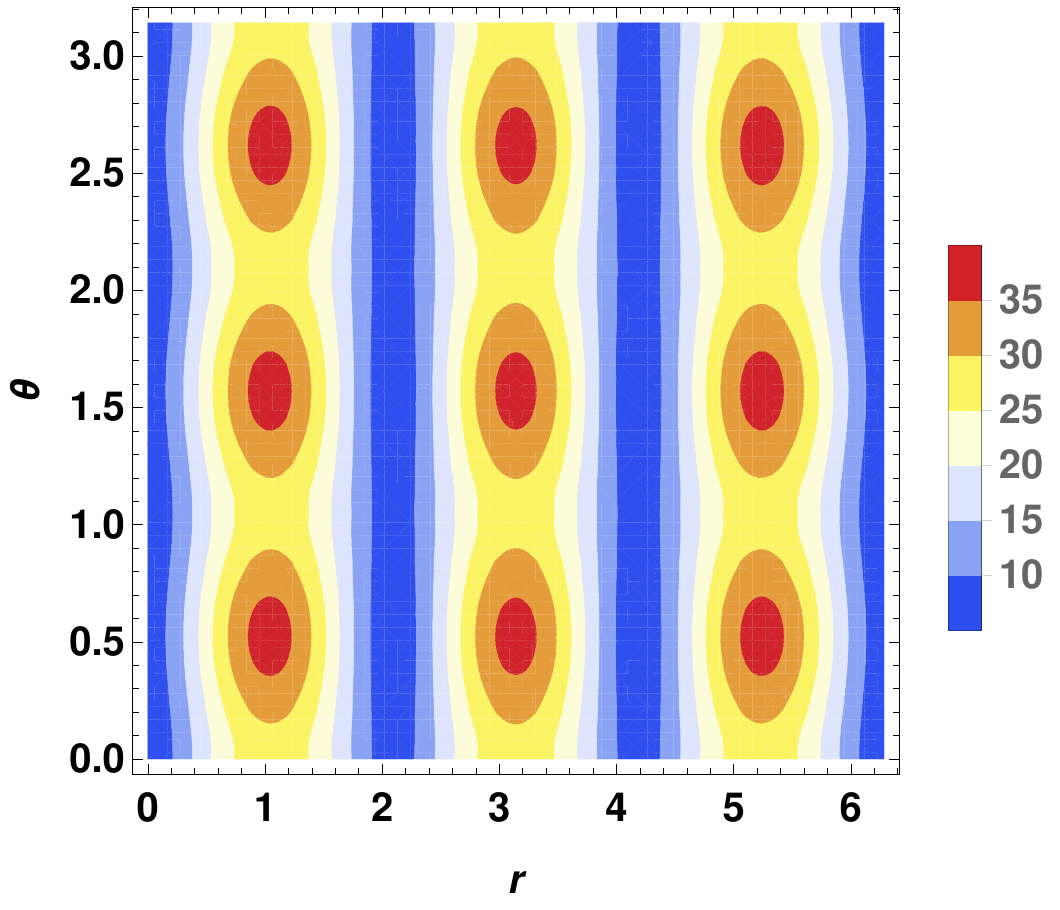}
\caption{Energy densities for Skyrmions without $\protect\omega$-mesons
(left column) and for Skyrmions with a cloud of $\protect\omega$-mesons
(rigth column) for $B=1$, $q=1$ and $B=3$, $q=3$ (top to bottom). The peaks
of the energy densities correspond with the peaks of the topological charge
density. It can be seen that the energy density is blurred in the presence of the $\omega$-mesons. Here the values of the couplings constants have been setting to $%
p=L=1$, $M_{\protect\pi}=M_{\protect\omega}=0$, $K=2$, $\protect\lambda=1$
and $g_{0}=0.17$. It is important to observe that the thightness of the spagheti is related to the box size and shape, as well as to $B/V$ where $B$ is the topological charge and $V$ is the volume of the box.}
\label{fig:LyAPlot}
\end{figure}

Fig. \ref{fig:LyAPlot} shows the energy density of the configurations. One
can see that, in fact, the system describes a lattice of baryonic tubes with
a crystalline order where the peaks of the
energy density are localized where the baryonic density takes its maximum
values and vanishes outside the tubes. 
In fact, the similarity of the contour plots with the spaghetti-like configurations found (numerically) in the
nuclear pasta phase in \cite{pasta1}, \cite{pasta2}, \cite{pasta2a}, \cite{pasta2b} and \cite{pasta3} is quite remarkable.

For this type of solutions the inclusion of the $\omega$-mesons in the Skyrme model causes 
the energy density distribution to blur due to the presence of this cloud of mesons, as can be seen comparatively in Fig. \ref{fig:LyAPlot}.
\begin{figure}[h!]
\centering
\includegraphics[scale=.7]{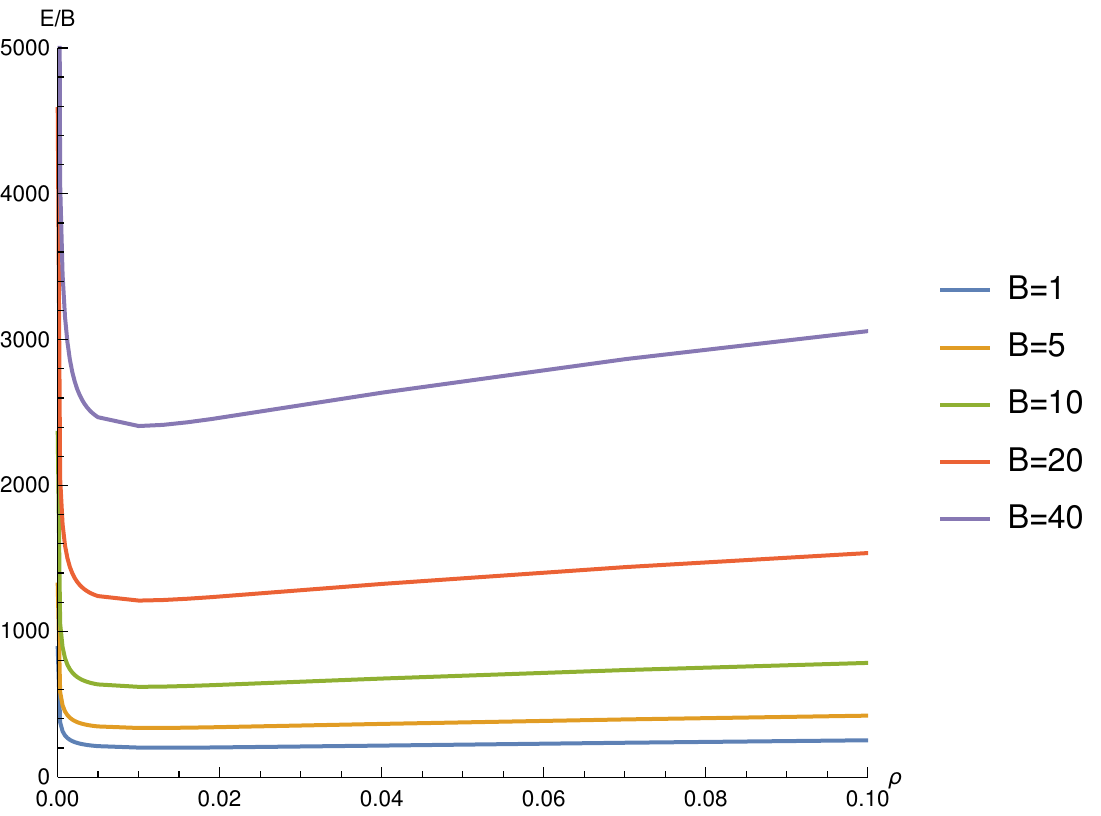}
\caption{Total energy per baryon as a function of the density for different
values of the baryonic charge of the Skyrmions with $\omega$-mesons.
The energy curves have the characteristic ``u-shape'' of nuclear pasta. Here
the coupling constants have been set as $K=2$, $\lambda=1$, $%
g_0=0.001$, $p=q=1$, $M_\protect\pi=M_\protect\omega=0$.}
\label{fig:ErhoOmega}
\end{figure}
In Fig. \ref{fig:ErhoOmega} one can see the total energy per baryon as a
function of the density, $\rho=1/V=1/(4\pi^3 L^3)$, of the box containing
the topological solitons. The energy curves have the characteristic
``u-shape'' of this kind of configurations: At low density the energy of the
system is a decreasing function of $\rho$, then there is a critical point
from which the behavior reverses in such a way that at higher densities the
energy of the system increases with $\rho$. In fact, Fig. \ref{fig:ErhoOmega}
is in qualitative agreement with what has been obtained in numerical simulations of
nuclear pasta (see \cite{pasta10}).

\begin{figure}[h]
\centering
\includegraphics[scale=1]{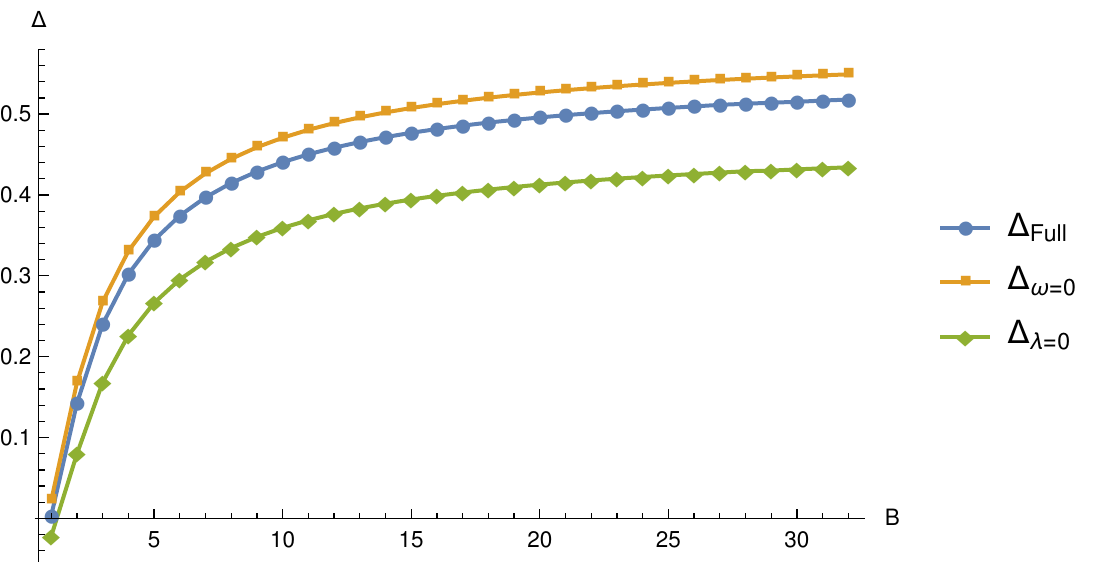}
\caption{$\Delta (B)$ for the following cases: The Skyrme theory coupled
with the $\protect\omega $-mesons ($\Delta _{\text{Full}}$), the Skyrme
theory without the $\protect\omega $-mesons ($\Delta _{\protect\omega =0}$)
and the NLSM coupled to the $\protect\omega $-mesons ($\Delta _{\protect%
\lambda =0}$). We can see that the inclusion of the $\protect\omega$-mesons
stabilize the Skyrmions. Here we have set $K=2$, $\protect\lambda =1$, $%
g_{0}=0.1$, $p=q=1$, $L=1$, $M_{\protect\pi }=M_{\protect\omega }=0$.}
\label{fig:DeltaOmega}
\end{figure}

Now we introduce an important quantity, $\Delta =\Delta (B)$, which is a
measure of the interaction energy between baryons 
\begin{equation}
\Delta (B)\ =\ \frac{E_{(B+1)}-(E_{(B)}+E_{(1)})}{\left( B+1\right) E_{(1)}}%
\ ,  \label{Delta}
\end{equation}%
where $B$ is the number of baryons of the system and $E_{(i)}$ the total
energy of a configuration containing $(i)$ baryons. As it is well known,
Skyrmions have a strong short range repulsion. This is reflected, for
instance, in the growth of the function $\Delta (B)$ with $B$ (see, for
instance, the plots in \cite{pasta10} of $E(B)/B$). On the other hand, due
to the stabilizing role of the $\omega$-mesons (see \cite{AN}), it is
natural to expect that the quantity $\Delta (B)$ when the $\omega $-mesons
are turned on ($\Delta _{\text{Full}}$) should grow, as function of $B$,
slower than when the $\omega $-mesons are turned off ($\Delta _{\omega =0}$%
). One of the main results of the present paper is that we can test this
intuition with exact analytic solutions of the Skyrme $\omega $-mesons
theory. In fact, the results in Fig. \ref{fig:DeltaOmega} confirm that for
any value of $B$ it is true that $\Delta _{\text{Full}}<\Delta _{\omega =0}$%
, therefore one can say that the $\omega$-mesons partially
\textquotedblleft shield" the baryon-baryon repulsion. The conclusion is
that it is crucial to take the effects of the $\omega $-mesons into account
in the analysis of complex configurations such as the ones of the nuclear
pasta phase. Also, from Fig. \ref{fig:DeltaOmega} we see the expected fact
that the stabilization of the Skyrmions is also achieved without the need to
consider the Skyrme term by coupling the baryonic current to the $\omega $%
-mesons (see the curve $\Delta _{\lambda =0}$). 


\section{Crystals of Baryons with a cloud of $\protect\rho$-mesons}


The Skyrme model produces nuclei binding energies larger than the
experimental ones. This problem can be solved by including the next lightest
Isospin 1 mesons, the $\rho$-mesons (see \cite{NS1} and \cite{NS2} and
references therein). At a first glance, the combined $\rho $-mesons Skyrme
theory able to fix the nuclei binding energy appears to be hopelessly
complicated to be studied with analytical methods. However, in this section
we will analyze relevant configurations of the family of actions introduced
in \cite{NS1} and \cite{NS2} in which a full analytic treatment is possible.

\subsection{The Skyrme $\protect\rho$-mesons theory}

The Skyrme $\rho $-mesons theory (see \cite{NS1} and \cite{NS2}) is
described by the action 
\begin{equation}
I[U,A_{\mu }]=\int \sqrt{-g} d^{4}x \biggl(\mathcal{L}_{\text{Skyrme}}+\mathcal{L}%
_{\rho }+\mathcal{L}_{\text{int}}\biggl)\ ,  \label{I2}
\end{equation}%
where 
\begin{equation*}
\mathcal{L}_{\text{Skyrme}}=\text{Tr}\left\{ \frac{c_{1}}{2}R_{\mu }R^{\mu }+%
\frac{c_{2}}{16}F_{\mu \nu }F^{\mu \nu }-\frac{M_{\pi }^{2}}{2}%
(U+U^{-1})\right\} \ ,
\end{equation*}%
corresponds to the usual Skyrme Lagrangian which includes the mass term for
the pions, and the tensors $R_{\mu }$ and $F_{\mu \nu }$ have been defined
in Section $2$ (see Eq. \eqref{I1}). Additionally, the Skyrme $\rho $-mesons
theory has the following two terms describing the $\rho $-mesons dynamics
and their interactions with the Skyrmions. These are 
\begin{align}
\mathcal{L}_{\rho }& =\text{Tr}\{a_{1}S_{\mu \nu }^{2}+a_{2}{}A_{\mu
}^{2}+c_{3}S_{\mu \nu }G^{\mu \nu }+c_{4}G_{\mu \nu }^{2}\}\ ,  \label{Lrho}
\\
\mathcal{L}_{\text{int}}& =\text{Tr}\{c_{5}L_{\mu \nu }^{2}-c_{6}F_{\mu \nu
}S^{\mu \nu }-c_{7}F_{\mu \nu }G^{\mu \nu }+a_{8}F_{\mu \nu }L^{\mu \nu
}+a_{9}L_{\mu \nu }S^{\mu \nu }+a_{10}L_{\mu \nu }G^{\mu \nu }\}\ ,
\label{Lint}
\end{align}%
where 
\begin{equation*}
G_{\mu \nu }=[A_{\mu },A_{\nu }]\ ,\quad L_{\mu \nu }=[R_{\mu },A_{\nu
}]-[R_{\nu },A_{\mu }]\ ,\quad S_{\mu \nu }=\partial _{\mu }A_{\nu
}-\partial _{\nu }A_{\mu }\ ,
\end{equation*}%
being $A_{\mu }$ the $SU(2)$-valued one-form that characterizes the $\rho $%
-mesons%
\begin{equation*}
A_{\mu }=\left( A_{\mu }\right) ^{a}t_{a}\ .
\end{equation*}%
In this section we have decided to denote as $a_{i}$ and $c_{i}$ the
coupling constants instead of $K$ and $\lambda $ in order to keep the value
of these as general as possible and also to compare our results with those
of the references \cite{NS1} and \cite{NS2}. In fact, we get to the action
considered in \cite{NS1} and \cite{NS2} if we set the $a_i$ couplings in
Eqs. \eqref{Lrho} and \eqref{Lint} as follows 
\begin{equation}
a_{1}=\frac{1}{8}c_{8}\ ,\ \ a_{2}=\frac{1}{4}m_{\rho }^{2}\ ,\ \ a_{8}=%
\frac{1}{2}c_{6}\ ,\ \ a_{9}=-\frac{1}{8}c_{8}\ ,\ \ a_{10}=-\frac{1}{2}%
c_{3}\ .  \label{toNS}
\end{equation}%
There are, at least, two sets of values for the constants $c_{i}$ that allow
to reduce the binding energy, namely 
\begin{gather}
c_{1}\rightarrow \frac{1}{4\sqrt{\pi }}\ ,\ c_{2}\rightarrow 0.198\ ,\
c_{3}\rightarrow \frac{1}{10}\ ,\ c_{4}\rightarrow \frac{1}{10}\ ,\
c_{5}\rightarrow 0.038\ ,\   \label{NS1} \\
c_{6}\rightarrow \frac{\pi ^{\frac{1}{4}}}{12\sqrt{2}}\ ,\ c_{7}\rightarrow
0.049\ ,\ c_{8}\rightarrow 1\ ,\ m_{\rho }\rightarrow \frac{1}{\sqrt{2}}\ , 
\notag
\end{gather}%
and 
\begin{gather}
c_{1}\rightarrow \frac{1}{6}\ ,\ c_{2}\rightarrow \frac{1}{6}\ ,\
c_{3}\rightarrow \frac{\sqrt{3}}{10}\ ,\ c_{4}\rightarrow \frac{9}{140}\ ,\
c_{5}\rightarrow \frac{3}{80}\ ,\   \label{NS2} \\
c_{6}\rightarrow \frac{\sqrt{3}}{24}\ ,\ c_{7}\rightarrow \frac{1}{20}\ ,\
c_{8}\rightarrow 1\ ,\ m_{\rho }\rightarrow \frac{2}{\sqrt{5}}\ .  \notag
\end{gather}%
However\footnote{The condition $c_8 = 1$ is required canonically because it is the unit residue of the propagator
at the $\rho$-meson mass.}, it is very likely that there are even more choices of the coupling
constants of the theory which allow to get good results for the binding
energies\footnote{%
We thanks C. Naya for this remark.}. Due to the large number of coupling
constants, it makes a lot of sense to explore such parameters space in order
to identify a smaller set of coupling constants which allows to simplify the
analysis without loosing the good features arising from the inclusion of the 
$\rho $-mesons. A good and effective criterion which will be adopted here is
to select the subset of couplings which allows a complete analytic solutions
of the field equations keeping, at the same time, good physical properties
(such as a reasonable behavior of the $\Delta (B)$\ defined in Eq. (\ref%
{Delta}) when the $\rho $-mesons are turned on). We will show that the
difference between our choices and the ones in \cite{NS1} and \cite{NS2} is
small.

The fifteen non-linear differential equations (see Appendix A for the
explicit derivation of these field equations) for the Skyrme $\rho$-mesons
are obtained by taking the variation of the action w.r.t the fields $U$ and $%
A_{\mu }$: 
\begin{align}
\nabla _{\mu }\biggl(c_{1}R^{\mu }+\frac{c_{2}}{4}[R_{\nu },F^{\mu \nu
}]-E_{1}^{\mu }\biggl)-[R_{\mu },E_{1}^{\mu }]-M_{\pi }^{2}(U-U^{-1})\quad
=& \quad 0\ ,  \label{EqU} \\
\nabla _{\nu }O_{1}^{\mu \nu }+2a_{2}A^{\mu }-E_{2}^{\mu }\quad =& \quad 0\ ,
\label{EqA}
\end{align}
where 
\begin{align}
E_{1}^{\mu }=\ & [S^{\mu \nu },2a_{9}A_{\nu }-2c_{6}R_{\nu }]+[G^{\mu \nu
},2a_{10}A_{\nu }-2c_{7}R_{\nu }]+2a_{8}[F^{\mu \nu },A_{\nu }]+[L^{\mu \nu
},4c_{5}A_{\nu }+2a_{8}R_{\nu }]\ ,  \label{Ef1} \\
E_{2}^{\mu }=\ & [S^{\mu \nu },2c_{3}A_{\nu }+2a_{9}R_{\nu }]+[G^{\mu \nu
},4c_{4}A_{\nu }+2a_{10}R_{\nu }]-[F^{\mu \nu },2c_{7}A_{\nu }-2a_{8}R_{\nu
}]  \notag \\
& +[L^{\mu \nu },2a_{10}A_{\nu }+4c_{5}R_{\nu }]\ ,  \label{Ef2} \\
O_{1}^{\mu \nu }=\ & 2\biggl(2a_{1}S^{\mu \nu }+c_{3}G^{\mu \nu
}-c_{6}F^{\mu \nu }+a_{9}L^{\mu \nu }\biggl)\ .
\end{align}

\subsection{The ansatz}

In order to constuct analytical solutions of the Skyrme $\rho$-mesons theory
we will use the same ansatz for the $U$ field presented in Eqs. \eqref{U1}, %
\eqref{U2} and \eqref{Uexplicit} together with a meron-like ansatz for the $%
\rho$-mesons field (see \cite%
{Alfaro}, \cite{meron0}, \cite{meron1}, \cite{meron2}, \cite{meron3} and 
\cite{meron4}) in the form: 
\begin{equation}  \label{meron}
\rho _{\mu }=\bar{\lambda} (r)U^{-1}\partial _{\mu }U\ ,
\end{equation}%
where $\bar{\lambda}$ will be called the ``profile" of the $\rho$-mesons. It
is worth to emphasize that, due to the fact that the $\rho$-mesons
Lagrangian is not gauge invariant one should not expect great
simplifications using an ansatz inspired from the Yang-Mills theory, however
we will show that the above ansatz works nevertheless. Here the metric it is
also the finite box defined in Eq. \eqref{Box}.

The energy density corresponding to the ansatz in Eqs. \eqref{Box}, %
\eqref{U1}, \eqref{U2}, \eqref{Uexplicit} and \eqref{meron} takes the form: 
\begin{equation}
\mathcal{E}=\mathcal{E}_{\text{Sk}}+\mathcal{E}_{\rho}+\mathcal{E}_{\text{int%
}} \ ,
\end{equation}%
where 
\begin{align}
\mathcal{E}_{\text{Sk}} =& \frac{c_{2} \sin ^2(\alpha) \left(2 p^2 \sin^2(q\theta  ) \left(q^2 \sin ^2(\alpha )+\alpha '^2\right)+q^2 \alpha '^2\right)}{L^4}+2 M_{\pi}^2 \cos(\alpha)\notag \\
&+\frac{c_{1} \left(\sin ^2(\alpha) \left(2 p^2 \sin^2(q\theta)+q^2\right)+\alpha '^2\right)}{L^2} \ , \\
\mathcal{E}_{\rho } =& \frac{\left.\sin ^2(\alpha) \left(p^2+q^2-p^2\sin(2 q\theta)\right) \left(\bar{\lambda}^2 \left(8 a_{1}+a_{2} L^2+8 \bar{\lambda}\alpha^{'2} (c_{4}\bar{\lambda}-c_{3})\right)+2 a_{1} \bar{\lambda}^{'2}\right)\right)}{L^4}\notag\\
&+\frac{16 p^2 q^2 \sin^2(q\theta) \sin ^4(\alpha) \bar{\lambda}^2\left(a_1-c_{3}\bar{\lambda}+c_4 \bar{\lambda}^2\right)+a_{2} L^2 \bar{\lambda}^2 \alpha^{'2}}{L^4} \ , \\
\mathcal{E}_{\text{int}} = & \frac{16 \bar{\lambda} \sin ^2(\alpha ) (\bar{\lambda} (2 a_{10} \bar{\lambda}-2 a_{9}+4 c_{5}-c_{7})+2 a_{8}+c_{6}) \left(2 p^2 \sin^2(q\theta) \left(q^2 \sin ^2(\alpha )+\alpha '^2\right)+q^2 \alpha '^2\right)}{L^4} \ .
\end{align}

\subsection{Analytical solutions with a constant meron profile}

A direct way to construct analytical solutions of the Skyrme $\rho $-mesons
theory is to assume a constant $\rho $-mesons profile\footnote{%
In Appendix B we will include a possible choice of the coupling constants
which allows an analytic solution with $\bar{\lambda}=\bar{\lambda}(r)$
non-constant.} , that is $\bar{\lambda}(r)=\lambda _{0}$, in the ansatz in
Eqs. \eqref{Box}, \eqref{U1}, \eqref{U2}, \eqref{Uexplicit} and \eqref{meron}
(in Yang-Mills terminology this would be a proper meron ansatz). Remarkably,
with this assumption the three Skyrme field equations can be reduced to just
one integrable equation for the soliton profile $\alpha (r)$ while the
twelve non-linear $\rho $-mesons field equations are reduced to three
polynomial equations for $\lambda _{0}$ provided that two extra constraints
on the coupling constants of the $\rho $-mesons action are satisfied.
Indeed, with $\bar{\lambda}(r)=\lambda _{0}$ and the $U$ field as in Eqs. (%
\ref{U1}), (\ref{U2}) and (\ref{Uexplicit}) the field equations read 
\begin{align}
a_{8}+\lambda _{0}\left( 3a_{10}\lambda _{0}-a_{9}-c_{3}\lambda
_{0}+2c_{4}\lambda _{0}^{2}+4c_{5}-c_{7}\right) & =0\ ,  \label{11} \\
c_{6}-\lambda _{0}(-2a_{1}+2a_{9}+c_{3}\lambda _{0})& =0\ ,  \label{12} \\
a_{2}& =0\ ,  \label{13} \\
\frac{\partial }{\partial r}\left[ \frac{1}{2}\bar{I}(\alpha )\alpha
^{\prime 2}-\bar{V}(\alpha )-\bar{E}_{0}\right] & =0\ ,  \label{aI}
\end{align}%
where $\bar{E}_{0}$ is an integration constant, and we have defined 
\begin{equation}
\bar{I}(\alpha )=2\tilde{a}_{1}q^{2}\sin ^{2}(\alpha )-2c_{1}L^{2}\ ,\qquad 
\bar{V}(\alpha )=\frac{1}{2}c_{1}L^{2}q^{2}\cos (2\alpha )+4L^{4}M_{\pi
}^{2}\cos (\alpha )\ ,  \label{IandV1}
\end{equation}%
together with 
\begin{equation*}
\tilde{a}_{1}=-\left(c_2+8\lambda_0\left(3a_8+c_6+\lambda_0\left(4c_{5}-c_{7}-a_{9}+a_{10}\lambda
_{0}\right)\right)\right) \ .
\end{equation*}%
First of all, it is worth to emphasize that the effects of the presence of
the $\rho $-mesons into the Skyrme field equations manifest themselves
through the effective coupling $\tilde{a}_{1}$ here above. Hence, in a
sense, the presence of the $\rho $-mesons manifests itself in a
renormalization of the coupling constants which would appear in the Skyrme
theory alone. The comparison between the equation for the profile $\alpha $
in Eqs. (\ref{Eqalpha}) and (\ref{YandV}) without $\rho $-mesons and the
Eqs. (\ref{aI}) and (\ref{IandV1}) that include the effects of the $\rho $%
-mesons is very instructive. It is quite amusing that in such a complicated
system it is possible to read off explicitly the effects of the $\rho $-mesons
on the Skyrmion profile $\alpha $. Secondly, this choice of the coupling (in
which the mass term of the $\rho $-mesons vanishes, $a_{2}=0$) is very
apropriate for large baryon number, in which case the mass of the $\rho $%
-mesons (as well as the mass of the pions $M_{\pi }$) can be neglected with
respect to the mass of the nuclear pasta configuration: obviously, the mass
of solitonic configurations with large baryonic charge is many orders of
magnitude larger than the $\rho $-mesons and pions masses (thus, in all the
plots here below we will assume that $a_{2}=M_{\pi }=0$).

Resuming, in order for the ansatz $\bar{\lambda}(r)=\lambda _{0}$ together
with the one in Eqs. (\ref{U1}), (\ref{U2}) and (\ref{Uexplicit}) to reduce
the complete set of fifteen coupled non-linear field equations to just one
integrable equation for $\alpha $ plus an algebraic equation for $\lambda
_{0}$, one only needs two constraints on the coupling constant of the $\rho $%
-mesons action: one constraint\footnote{%
The larger is the baryonic charge of the solitonic configuration, the more
accurate approximation $a_{2}=0$ becomes. Thus, in the present context we
can safely assume it.} is $a_{2}=0$ while the second constraint on the
coupling constant can be deduced by solving Eq. (\ref{12}) for $\lambda _{0}$
and then replacing this solution into Eq. (\ref{11}). We will discuss such a
constraint in the next subsections. 
\begin{figure}[h]
\centering
\includegraphics[scale=.9]{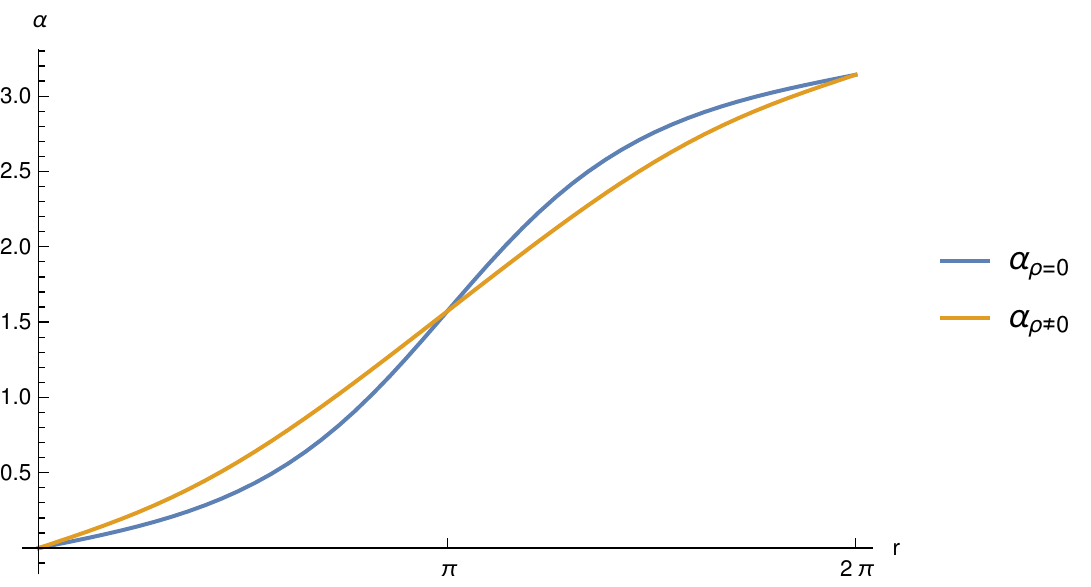}
\caption{Comparison between the profile of the Skyrmion with and without the
contribution of the $\protect\rho $-mesons for $B=1$. We see that the
inclusion of the $\protect\rho $-mesons smooths out the behavior of the soliton
profile.}
\label{fig:profrho}
\end{figure}
Also in this case it is possible to integrate the equation for the soliton
profile in Eq. \eqref{aI} analytically as follows 
\begin{equation}
\frac{d\alpha }{\eta _{\rho }(\alpha ,\bar{E}_{0})}=\pm dr\ ,\quad \eta
_{\rho }(\alpha ,\bar{E}_{0})=\frac{[2(\bar{E}_{0}+\bar{V}(\alpha ))]^{1/2}}{%
\bar{I}(\alpha )^{1/2}}\ ,  \label{alpharho}
\end{equation}%
where $\bar{E}_{0}$ is fixed by the boundary conditions, namely 
\begin{equation*}
\int_{0}^{n\pi }\frac{d\alpha }{\eta _{\rho }\left( \alpha ,\bar{E}%
_{0}\right) }=n\int_{0}^{\pi }\frac{d\alpha }{\eta _{\rho }\left( \alpha ,%
\bar{E}_{0}\right) }=2\pi \ .
\end{equation*}

\subsection{Binding energy on the crystal}

As it has been already remarked, there are two remaining equations, Eqs. %
\eqref{11} and \eqref{12}, which will fix $\lambda _{0}$ in terms of the
couplings of the theory and will give an extra constraint on the couplings.
A bound on the coupling constants arises by requiring that the solution $%
\lambda _{0}$ should be real, so the discriminant of Eq. \eqref{12} has to
be positive, that is 
\begin{equation}
a_{1}^{2}-2a_{1}a_{9}+a_{9}^{2}+c_{3}c_{6}>0\ .  \label{constrour}
\end{equation}%
Interestingly enough, \textit{this inequality is satisfied by both set of
parameters in} Eqs. \eqref{NS1} and \eqref{NS2} (see \cite{NS1} and \cite{NS2}). It is worth emphasizing
here that the above constraint in Eq. (\ref{constrour}) on the coupling
constants of the $\rho $-mesons action arises \textit{%
if one insists} that our approach must reduce consistently the complete set
of field equations of the Skyrme $\rho$-mesons theory to just one
integrable ODE for $\alpha $ plus an algebraic equation for $\lambda _{0}$.
On the other hand, the two choices of coupling constants analyzed in \cite%
{NS1} and \cite{NS2} arise with the physically well-motivated requirement to
reduce the nuclei binding energies of the Skyrme model. Obviously, there is, 
\textit{a priori}, no reason why one should expect that the constraint in
Eq. (\ref{constrour}) (which is a necessary condition in order to ensure
that our ansatz works) should be satisfied by the two choices of coupling
constants in \cite{NS1} and \cite{NS2} (motivated by the need to improve the
prediction of the nuclei binding energy of the Skyrme model). Nevertheless,
the constraint in Eq. (\ref{constrour}) is satisfied by both choices of
coupling constants in \cite{NS1} and \cite{NS2} (see Eqs. \eqref{NS1} and \eqref{NS2}).

From Eqs. \eqref{11} and \eqref{12} we can
choose the coupling $c_{4}$ as the \textquotedblleft dependent one": namely,
we will solve $c_{4}$ in terms of the other coupling constants while the
values of the other coupling constants will be as in Eqs. \eqref{NS1} and \eqref{NS2}. In Table \ref{Tabc4} we show the four values of the dependent coupling $c_4$ that solves the constraints. 

\begin{table}[h]
\begin{center}
\begin{tabular}{|c||c|c|}
\hline
$c_4$ & $\lambda_0$ positive root & $\lambda_0$ negative root \\ \hline\hline
Eq. \eqref{NS1} & $0.0198233$ & $-0.182293$ \\ \hline
Eq. \eqref{NS2} & $0.0586335$ & $-0.598634$ \\ \hline
\end{tabular}%
\end{center}
\caption{The four values of $c_{4}$ that solves the constaints in Eqs. \eqref{11} and \eqref{12}, according to the coupling constants in Eqs. \eqref{NS1} and \eqref{NS2}.}
\label{Tabc4}
\end{table}

The differences between the values of $c_{4}$ taking into account
Eqs. \eqref{11} and \eqref{12} and the set of values as in Eqs. \eqref{NS1} and \eqref{NS2} are shown in Table \ref{Tab}.

\begin{table}[h]
\begin{center}
\begin{tabular}{|c||c|c|}
\hline
& $\lambda_0$ positive root & $\lambda_0$ negative root \\ \hline\hline
$|c_{4}^{(1)}-c_{4}|/c_{4}^{(1)}$ & $0.801767$ & $2.82293$ \\ \hline
$|c_{4}^{(2)}-c_{4}|/c_{4}^{(2)}$ & $0.0879228$ & $10.3121$ \\ \hline
\end{tabular}%
\end{center}
\caption{Difference between the coupling constant $c_{4}$ in Eqs. \eqref{NS1} and \eqref{NS2} (found in
references \protect\cite{NS1} and \protect\cite{NS2}) and the two values of $c_{4}$ that came from the
two branches that solve Eqs. \eqref{11} and \eqref{12}. One can see that the
positive root is extremely close to one of the values that reduces the
binding energy.}
\label{Tab}
\end{table}
One can see that the positive root for $c_{4}$ is very close to the choice
in Eq. \eqref{NS2}. This is a quite non-trivial result
since there is no obvious relation between the condition that ``\textit{the
present ansatz should work}" for the Skyrme $\rho$-mesons theory and the
physically well-motivated condition in references \cite{NS1} and \cite{NS2} to reduce
the nuclei binding energies. Notwithstanding this, among the infinitely many
values that $c_{4}$ could have, the mathematical consistency of the present
approach produces a value for $c_{4}$ which is very close to the one in Eq. \eqref{NS2}. These results clearly show that the present analytic
approach is very well suited to describe multi-solitonic solutions of the
Skyrme-vector mesons theory. 
\begin{figure}[h]
\centering
\includegraphics[scale=0.42]{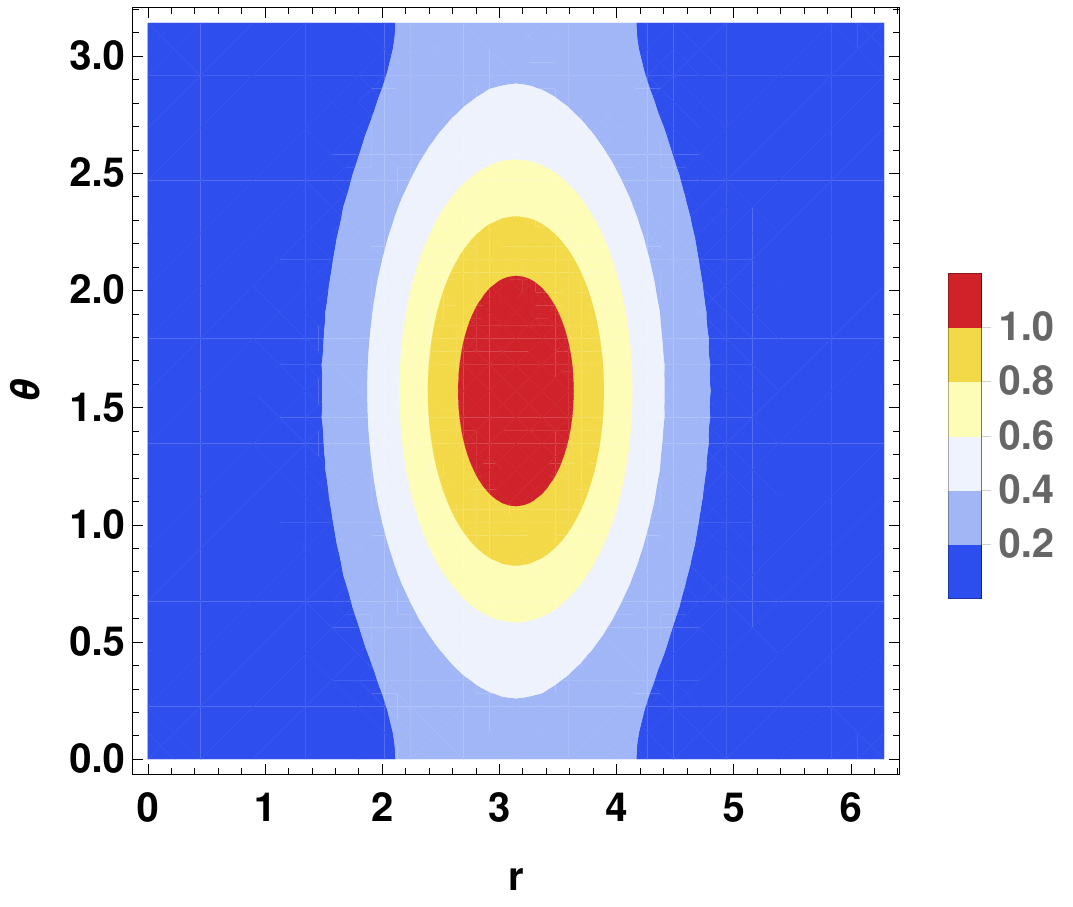} \qquad %
\includegraphics[scale=0.42]{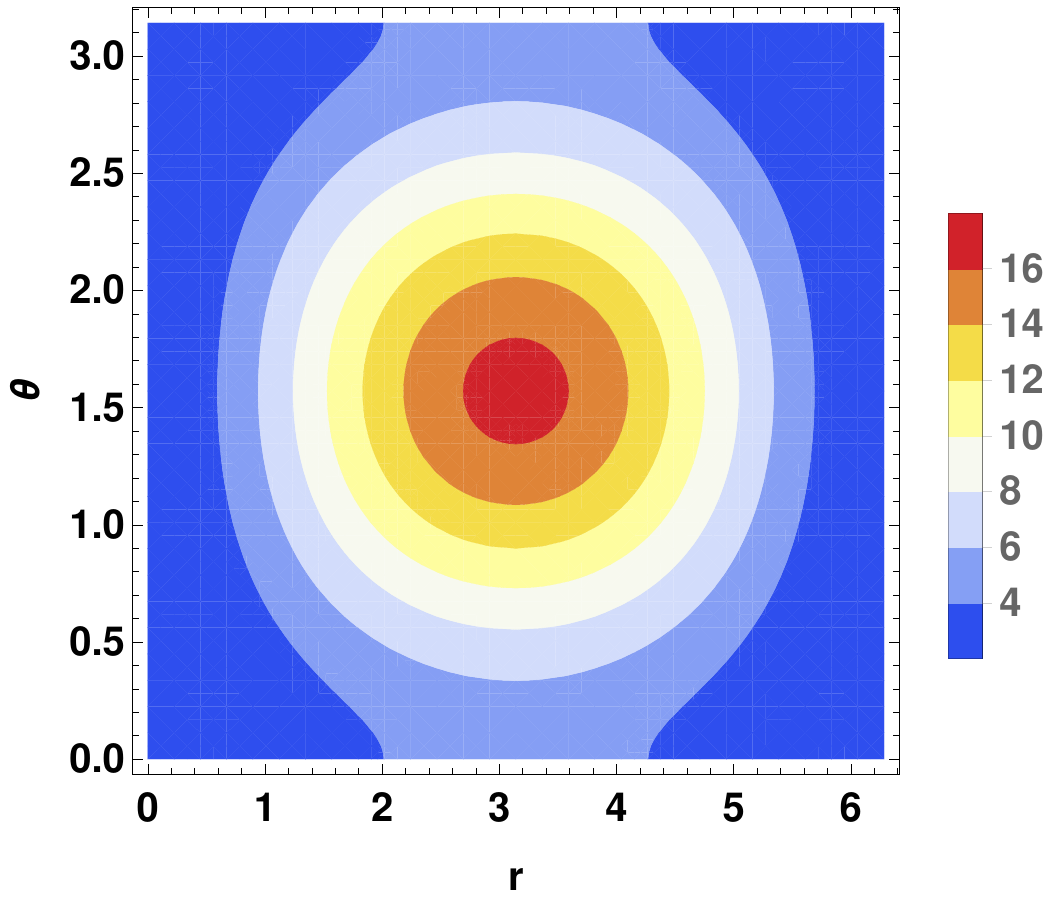} \qquad \qquad \qquad \qquad \qquad %
\includegraphics[scale=0.42]{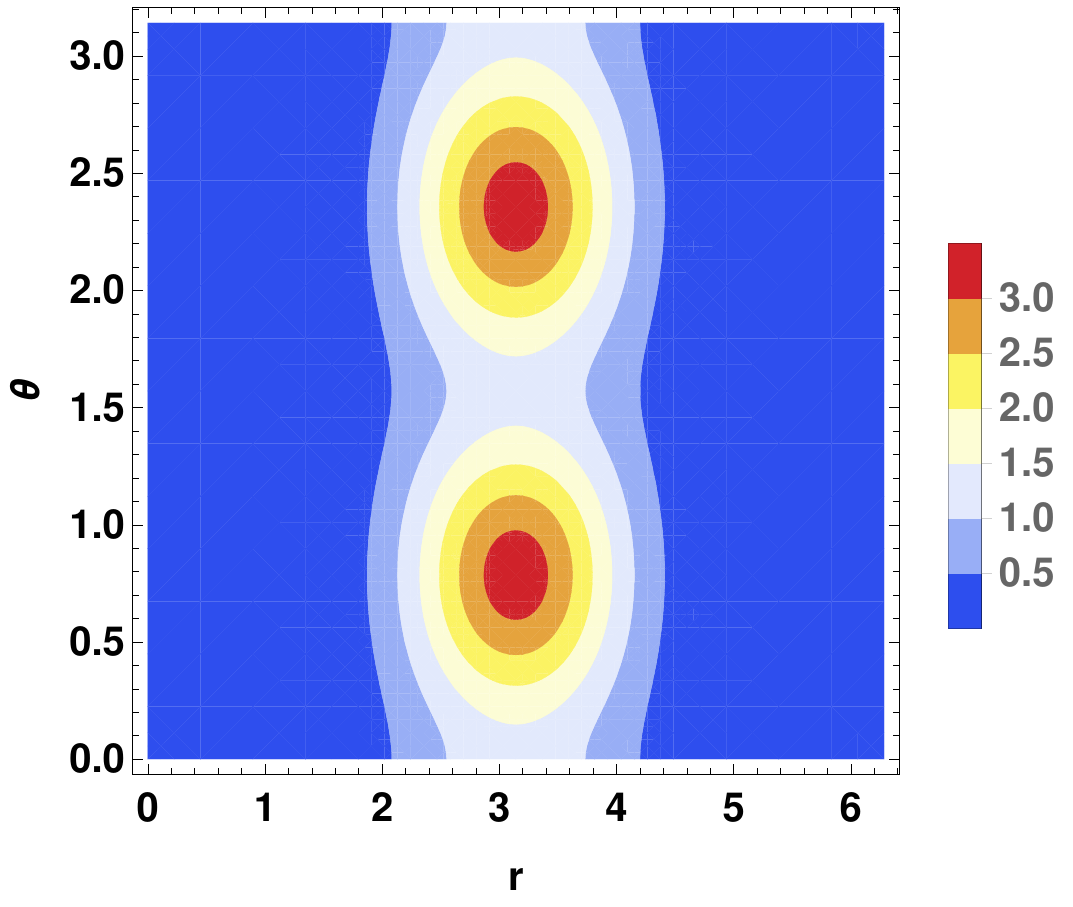} \qquad %
\includegraphics[scale=0.42]{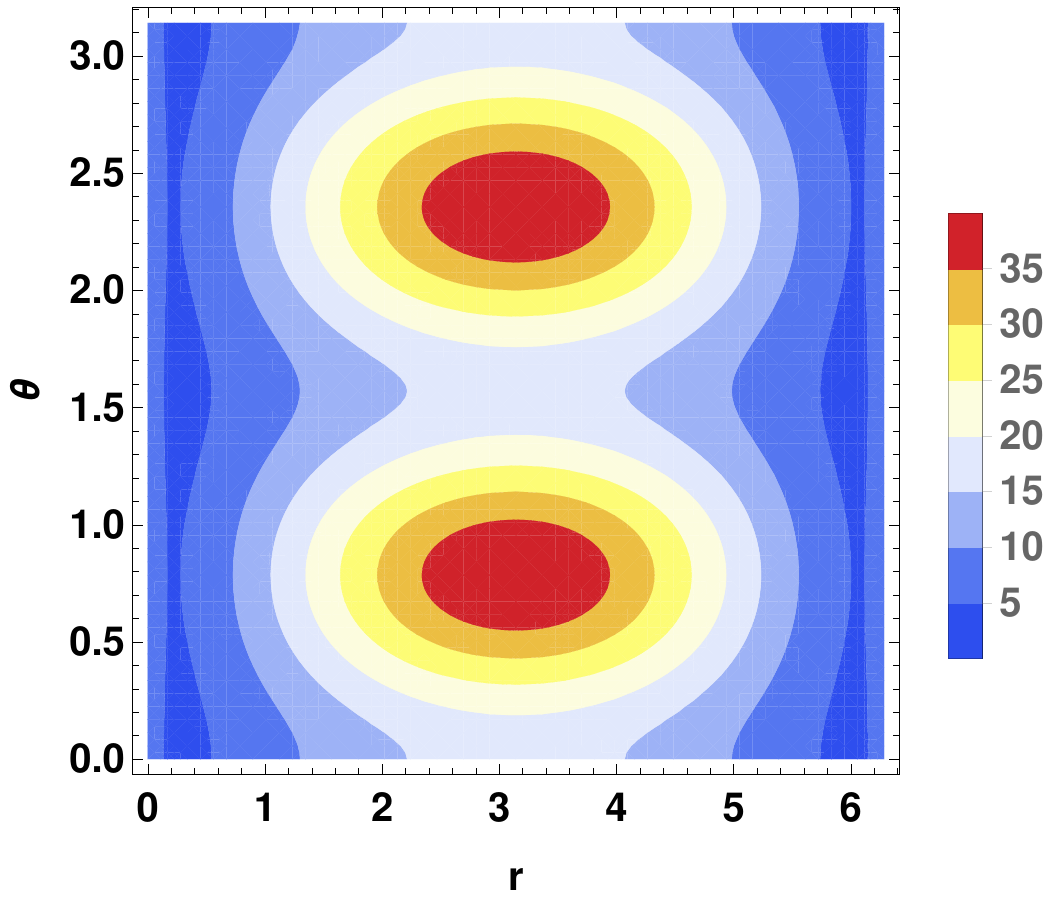}
\caption{Energy densities for Skyrmions without $\rho$-mesons (left
column) and for Skyrmions with a cloud of $\rho $-mesons (rigth
column) for $B=1$, $q=1$ and $B=2$, $q=1$ (top to bottom). The peaks of the
energy densities correspond with the peaks of the topological charge
density. The presence of the $\rho$-mesons turns the transverse sections of the tubes into a circular shape. Here the values of the coupling constants have been setting to $%
p=L=1$, $M_{\protect\pi }=a_{2}=0$, $c_{4}=-0.5986$ and the remaining $c_{i}$
are fixed as in Eq. \eqref{NS2}.}
\label{rhodensity}
\end{figure}

Fig. \ref{rhodensity} shows the energy density for Skyrmionic configurations with and without $\rho$-mesons.
In the absence of $\rho$-mesons the tubes of Skyrmions have the shape of ellipses (even when the theory is coupled to $\omega$-mesons, as we showed in the previous section), however, here we can see that this characteristic is modified due to the cloud of $\rho$-mesons. In fact, the presence of these vector mesons turns the transverse sections of the tubes into a circular shape (in the $r-\theta$ plane).

\begin{figure}[h!]
\centering
\includegraphics[scale=.8]{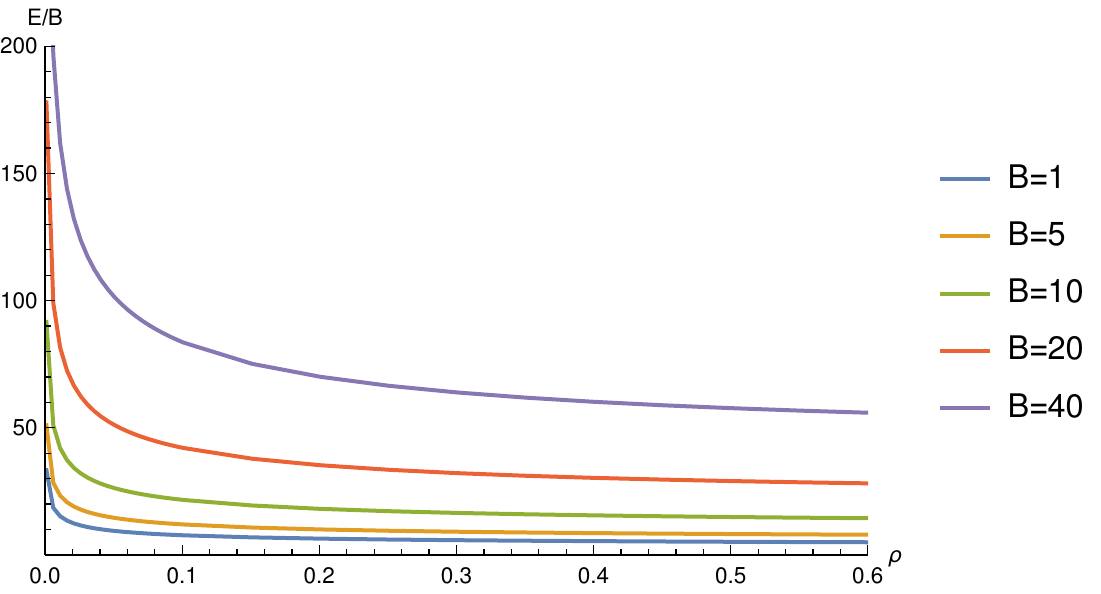}
\caption{Total energy per baryon as a function of the density for different
values of the baryonic charge for the Skyrmions with $\protect\rho$-mesons.
The coupling constants have been set as $p=q=1$, $M_\protect\pi=a_2=0$, $%
c_4=-0.1823$ and the remaining $c_i$ are fixed as in Eq. \eqref{NS1}.}
\label{fig:Erhorho}
\end{figure}

In Fig. \ref{fig:Erhorho} we show the energy of the solitons coupled with $%
\rho $-mesons in the low density sector, where the curves for different
values of the topological charge can be clearly distinguished. In the high
density sector the energy becomes an increasing function in such a way that
the energy (per baryon) versus density has the characteristic
\textquotedblleft u-shape\textquotedblright\ of nuclear pasta. A relevant
fact that can be seen from the energy plots is that, for the range of
parameters that we have chosen here, the presence of the $\rho $-mesons does
not spoil the good properties of the Skyrme model: in particular, the energy
conditions are satisfied in agreement with Wald theorem.

\begin{figure}[h!]
\centering
\includegraphics[scale=.7]{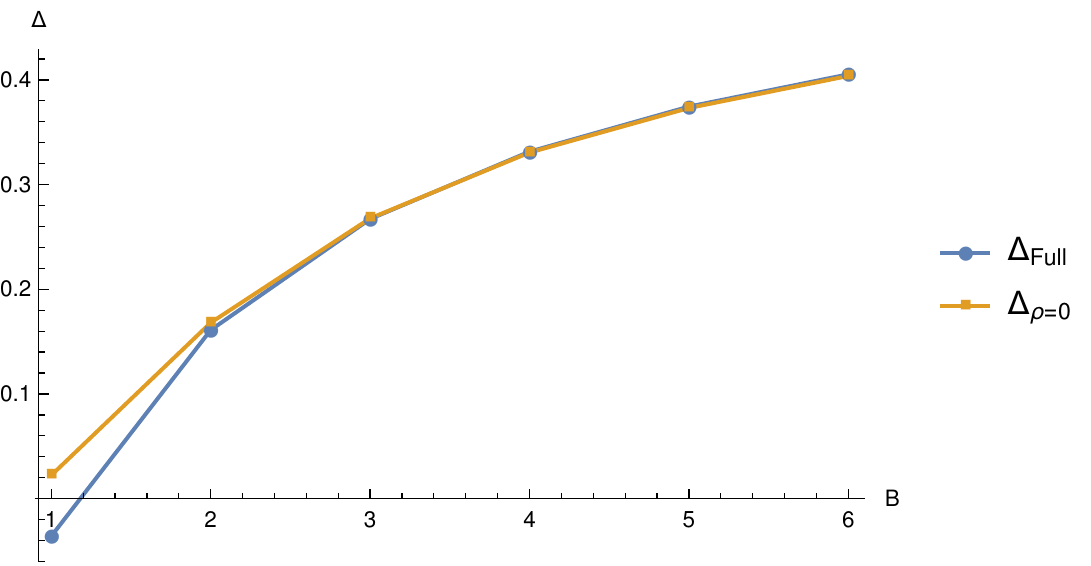}
\caption{$\Delta(B)$ for the following cases: The Skyrme theory coupled with
the $\rho$-mesons ($\Delta_{\text{Full}}$) with the solution $c_4=0.0586335$ and the Skyrme theory
without the $\rho$-mesons ($\Delta_{\protect\rho=0}$). The inclusion of the $\rho$-mesons
reduces the binding energy between the Skyrmions for low values of $B$.}
\label{fig:DeltaRho1}
\end{figure}

\begin{figure}[h!]
\centering
\includegraphics[scale=.35]{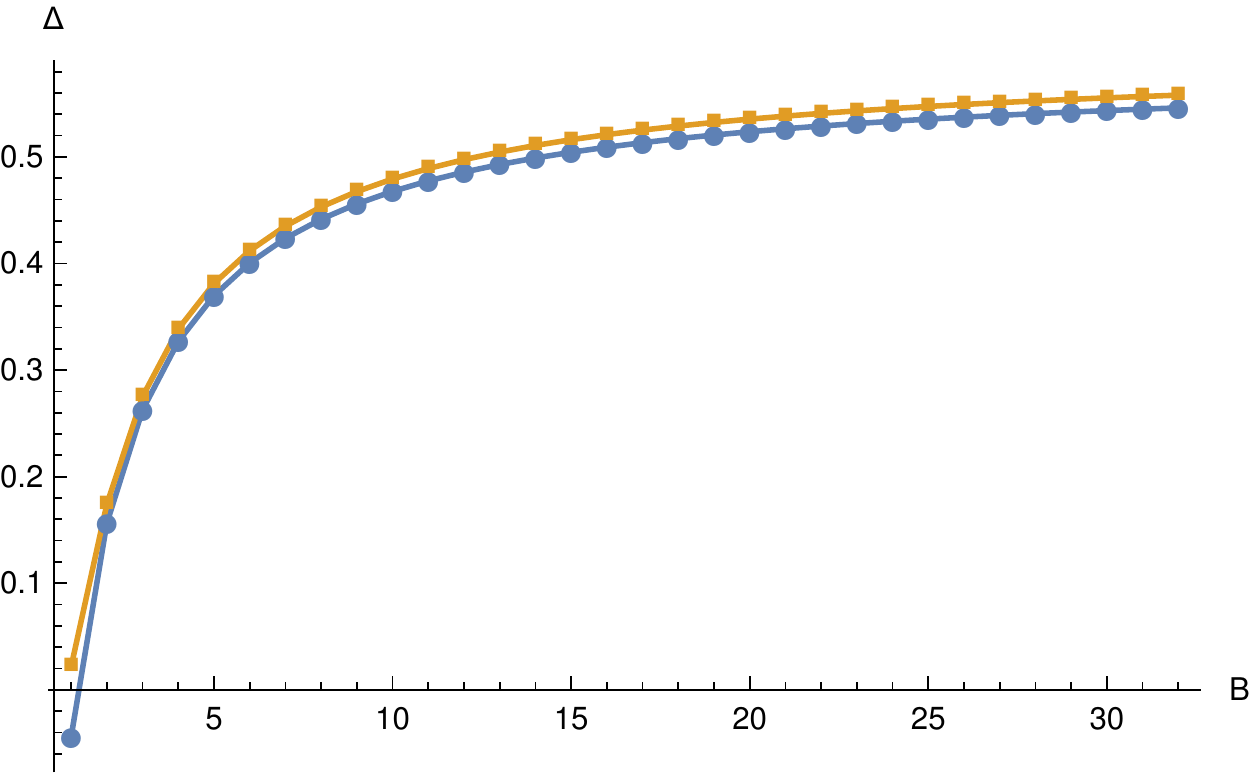}\qquad
\includegraphics[scale=.35]{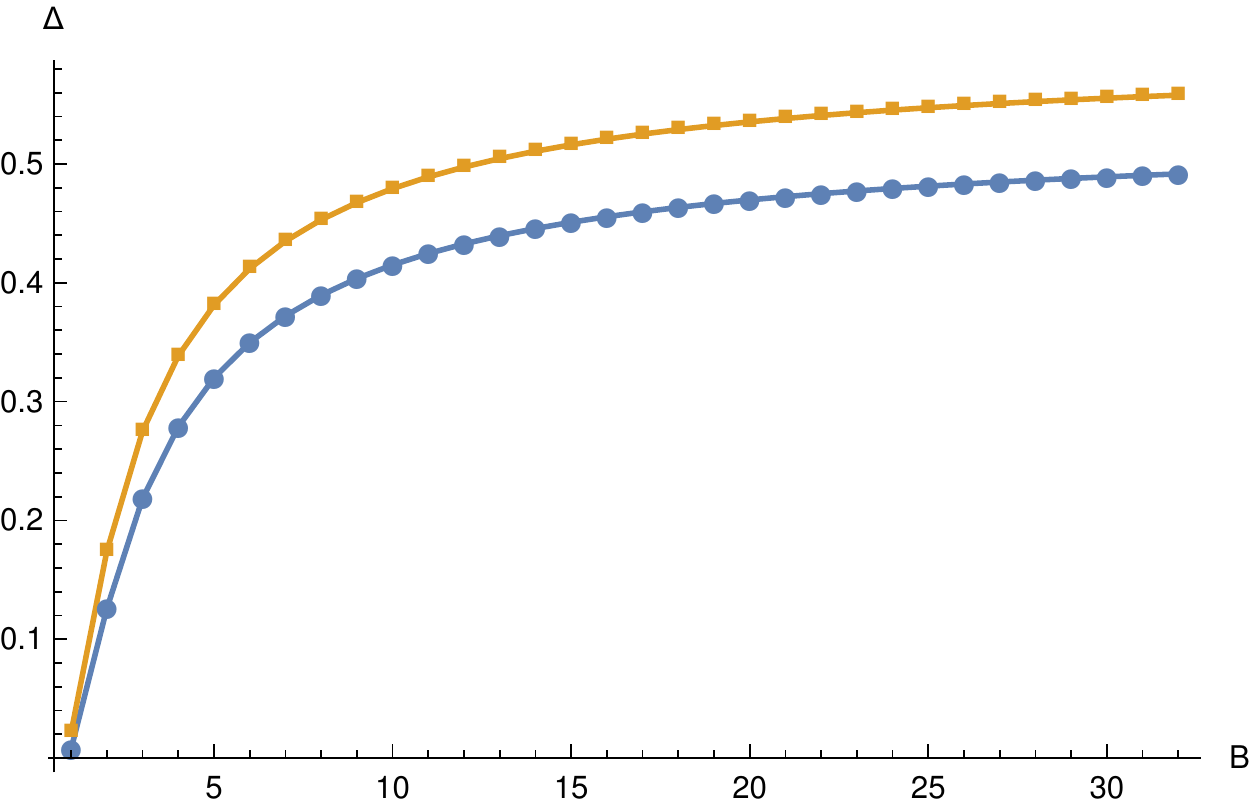}\qquad
\includegraphics[scale=.48]{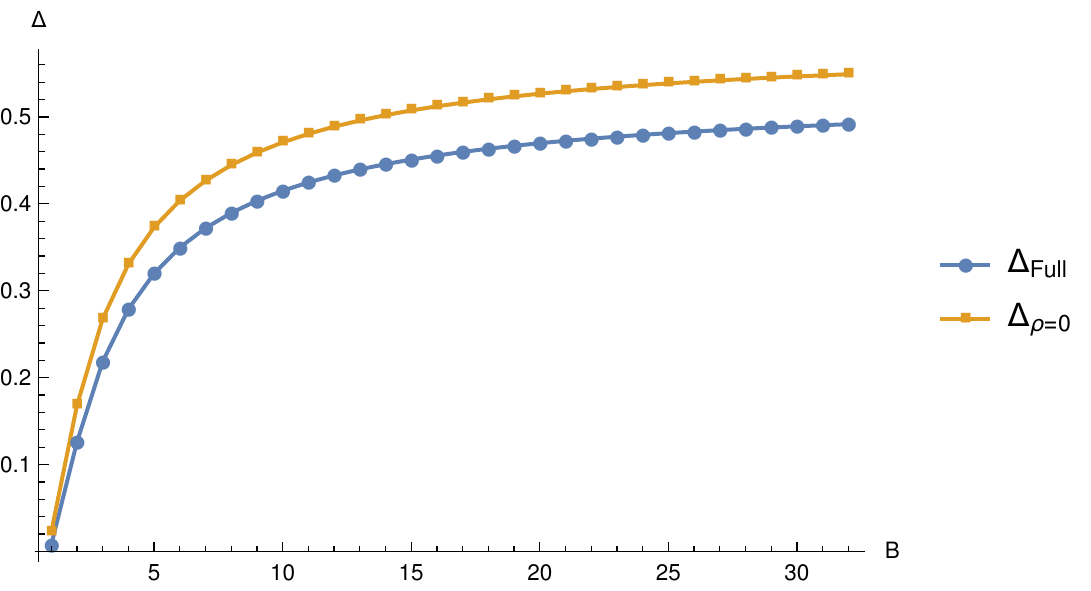}
\caption{$\Delta(B)$ for the following cases: The Skyrme theory coupled with
the $\rho$-mesons ($\Delta_{\text{Full}}$) with $c_4=0.0198233$, $-0.182293$ and $-0.598634$ (from left to right) and the Skyrme theory
without the $\rho$-mesons ($\Delta_{\rho=0}$). The inclusion of the $\rho$-mesons
reduces the binding energy between the Skyrmions at every value of the baryonic charge.}
\label{fig:DeltaRho2}
\end{figure}

Finally, we show a measure of the interaction energy between baryons using Eq. \eqref{Delta}. In Fig. \ref{fig:DeltaRho1} it is shown the case with $c_4$ being the closest value to the results of \cite{NS1} and \cite{NS2}. We can see that the inclusion of the $\rho$-mesons reduces the binding energy between the Skyrmions for small values of baryonic charge. Above a critical value of $B$, the behavior changes and the Skyrmionic binding energy increases due to the presence of the $\rho$-mesons. In Fig. \ref{fig:DeltaRho2} we have $\Delta$ for the remaining allowed values of $c_4$. Here it is clear that the inclusion of the $\rho$-meson reduces the binding energy between the Skyrmions, for all the values of the baryonic charge.
These cases are very relevant since it is very likely that there exist more good choices of the coupling constants of the theory, besides the ones found in \cite{NS1} and \cite{NS2}, as we already mentioned above.

Finally, in order to respect chiral symmetry one can add axial chiral mesons to the theory as,
in references \cite{Kaymakcalan} and \cite{Bando}, or to eliminate the coupling constant $a_1$ in Eq. \eqref{Lrho}.
In both cases, our asantz still solves the field equations,
and the soliton configurations maintain their physical properties. 


\section{Conclusions}


In the present manuscript we have constructed the first analytic examples of
hadronic crystals at finite baryon density for the Skyrme $\omega $-mesons
model and for the Skyrme $\rho $-mesons theory. These multi-Skyrmionic
configurations represent crystals of baryonic tubes surrounded by a cloud of
vector -mesons at finite baryon density. For the Skyrme $\omega$-mesons
case a suitable ansatz for the Skyrmions and the $\omega$-mesons reduce
consistently the complete set of seven coupled non-linear field equations to
just two integrable differential equations, one for the Skyrme profile and
the other for the $\omega$-mesons (which is actually a two-dimensional
Poisson equation in which the Skyrmion profile acts as a source term). This
analytical construction allows to show explicitly how the inclusion of $%
\omega $-mesons in the Skyrme action reduces the repulsive interaction
energy between baryons. It is possible to analyze explicitly the effects of
the $\omega $-mesons on various observables relevant for the nuclear pasta
phase: the present results strongly suggest that the $\omega$-mesons can be
important in the nuclear spaghetti phase. Using a similar approach, it is
also possible to include the $\rho $-mesons. A priori, the Skyrme $\rho$%
-mesons system introduced in \cite{NS1} and \cite{NS2} is even more
complicated than the Skyrme $\omega$-mesons system since the $\rho$-mesons
are described as a one-form taking values in the Lie algebra of $SU(2)$ and,
moreover, the family of actions in \cite{NS1} and \cite{NS2} contains many
very complicated interaction terms. Nevertheless, if one insists in using
the same ansatz for the Skyrme field together with a meron-like ansatz for
the $\rho$-mesons, one can reduce the complete set of fifteen non-linear
field equations to just one integrable equation for the Skyrmion profile and
an algebraic equation for the $\rho$-mesons profile provided a suitable
relation between the coupling constants of the $\rho$-mesons theory holds.
Quite remarkably, if one expresses one of the coupling constant (say, $c_{4}$%
) in terms of the others (using as input the values in \cite{NS1} and \cite%
{NS2}) the constraint that arises requiring the consistency of the ansatz
when the $\rho$-mesons are included provides a value for $c_{4}$ very close
to the ones in \cite{NS1} and \cite{NS2}. These analytic results on the
Skyrme $\rho$-mesons system show that a proper analytic description of the
nuclear pasta phase should include these vector-mesons.

\subsection*{Acknowledgements}

The authors are very grateful to Julio Oliva for his useful comments and
collaboration in the first stage of this project. The authors would also
like to thanks Carlos Naya for very useful suggestions and comments. F. C.
has been funded by Fondecyt Grant No. 1200022. M.L. is funded by FONDECYT
post-doctoral Grant No. 3190873. A.V. is funded by FONDECYT post-doctoral Grant No.
3200884. G. B. has been partially funded by FONDECYT Grant No. 1181047. The
Centro de Estudios Cient\'{\i}ficos (CECs) is funded by the Chilean
Government through the Centers of Excellence Base Financing Program of ANID.

\section*{Appendix A: Field equations for the Skyrme $\protect\rho$-mesons
theory}

\label{App1}

The variation of the Lagrangian defined in Eq. (\ref{I1}) w.r.t. the
fundamental fields $U$ and $A_{\mu}$ leads to 
\begin{align*}
\delta\mathcal{L}= & \text{Tr} \left\lbrace c_{1}\delta R_{\mu} R^{\mu }+%
\frac{c_{2}}{8}\delta F_{\mu\nu}F^{\mu\nu}-\frac{m_{\pi}c_{1}^{2}}{2c_{2}}%
(\delta U+\delta U^{-1})\right\rbrace \\
& +\text{Tr}\left\lbrace 2a_{2}\delta A_{\mu}A^{\mu}+\frac{1}{2}(\delta
S_{\mu\nu}O_{1}^{\mu\nu}+\delta G_{\mu\nu}O_{2}^{\mu\nu}+\delta L_{\mu\nu
}O_{3}^{\mu\nu}+\delta F_{\mu\nu}O_{4}^{\mu\nu}) \right\rbrace \ ,
\end{align*}
where we have defined the antisymmetric tensors $O_{i}^{\mu\nu}$ as follow: 
\begin{align*}
\frac{1}{2}O_{1}^{\mu\nu} & =2a_{1}S^{\mu\nu}+c_{3}G^{\mu\nu}-c_{6}F^{\mu
\nu}+a_{9}L^{\mu\nu} \ , \\
\frac{1}{2}O_{2}^{\mu\nu} & =c_{3}S^{\mu\nu}+2c_{4}G^{\mu\nu}-c_{7}F^{\mu
\nu}+a_{10}L^{\mu\nu} \ , \\
\frac{1}{2}O_{3}^{\mu\nu} & =a_{9}S^{\mu\nu}+a_{10}G^{\mu\nu}+a_{8}F^{\mu
\nu}+2c_{5}L^{\mu\nu} \ , \\
\frac{1}{2}O_{4}^{\mu\nu} &
=-c_{6}S^{\mu\nu}-c_{7}G^{\mu\nu}+a_{8}L^{\mu\nu} \ .
\end{align*}
It is direct to verify that 
\begin{align*}
\text{Tr}(\delta S_{\mu\nu}O_{1}^{\mu\nu})= & 2\text{Tr}(\delta
A_{\mu}\nabla_{\nu}O_{1}^{\mu\nu}) \ , \\
\text{Tr}(\delta G_{\mu\nu}O_{2}^{\mu\nu})= & 2\text{Tr}(\delta
A_{\mu}[A_{\nu},O_{2}^{\mu\nu}]) \ , \\
\text{Tr}(\delta L_{\mu\nu}O_{3}^{\mu\nu})= & 2\text{Tr}(\delta
R_{\mu}[A_{\nu},O_{3}^{\mu\nu }]-\delta A_{\mu}[O_{3}^{\mu\nu},R_{\nu}]) \ ,
\\
\text{Tr}(\delta F_{\mu\nu}O_{4}^{\mu\nu})= & 2\text{Tr}(\delta
R_{\mu}[R_{\nu},O_{4}^{\mu\nu}]) \ .
\end{align*}
Now, replacing the above in the variation of the Lagrangian and grouping
terms we obtain 
\begin{align*}
\delta\mathcal{L}= & \text{Tr} \left\lbrace \delta R_{\mu}\left(
c_{1}R^{\mu}+\frac{c_{2}}{4}[R_{\nu},F^{\mu\nu}]-[O_{3}^{\mu\nu},A_{\nu
}]-[O_{4}^{\mu\nu},R_{\nu}] \right) -\frac{m_{\pi}c_{1}^{2}}{2c_{2}}(\delta
U+ \delta U^{-1}) \right\rbrace \\
& + \text{Tr} \left\lbrace \delta A_{\mu} \left( \nabla_{\nu}O_{1}^{\mu\nu
}+2a_{2}A^{\mu}-[O_{2}^{\mu\nu},A_{\nu}]-[O_{3}^{\mu\nu},R_{\nu}] \right)
\right\rbrace \ .
\end{align*}
Taking into account that 
\begin{align*}
\text{Tr}(\delta R_{\mu}C^{\mu})=\text{Tr}(U^{-1}\delta
U([C^{\mu},R_{\mu}]-\nabla_{\mu}C^{\mu})) \ ,
\end{align*}
for an arbitrary tensor $C^{\mu}$, it follows that 
\begin{align*}
\delta\mathcal{L}= & \text{Tr} \left\lbrace U^{-1}\delta U\left(
-c_{1}\nabla_{\mu}R^{\mu }-\frac{c_{2}}{4}\nabla_{\mu}[R_{\nu},F^{\mu\nu}]-%
\frac{m_{\pi}^{2}c_{1}^{2}}{2c_{2}}(U-U^{-1})+\nabla_{\mu}E_{1}^{%
\mu}-[E_{1}^{\mu},R_{\mu}]\right) \right\rbrace \\
& +\text{Tr} \left\lbrace \delta A_{\mu} \left( \nabla_{\nu}O_{1}^{\mu\nu
}+2a_{2}A^{\mu}-E_{2}^{\mu} \right) \right\rbrace \ ,
\end{align*}
where $E_{i}^{\mu}$ have been defined as 
\begin{align*}
E_{1}^{\mu}= & [O_{3}^{\mu\nu},A_{\mu}]+[O_{4}^{\mu\nu},R_{\nu}] \\
= & [S^{\mu\nu},2a_{9}A_{\nu}-2c_{6}R_{\nu}]+[G^{\mu\nu},2a_{10}A_{\nu
}-2c_{7}R_{\nu}]+2a_{8}[F^{\mu\nu},A_{\nu}]+[L^{\mu\nu},4c_{5}A_{%
\nu}+2a_{8}R_{\nu}] \ ,  \label{Eff1} \\
E_{2}^{\mu}= & [O_{2}^{\mu\nu},A_{\nu}]+[O_{3}^{\mu\nu},R_{\nu}] \\
= & [S^{\mu\nu},2c_{3}A_{\nu}+2a_{9}R_{\nu}]+[G^{\mu\nu},4c_{4}A_{\nu
}+2a_{10}R_{\nu}] +[F^{\mu\nu},-2c_{7}A_{\nu}+2a_{8}R_{\nu}]+[L^{\mu\nu
},2a_{10}A_{\nu}+4c_{5}R_{\nu}] \ .
\end{align*}
The energy-momentum tensor of the theory reads 
\begin{align*}
T_{\mu\nu}=T^{\text{Sk}}_{\mu\nu}+T^{\rho}_{\mu\nu}+T^{\text{int}}_{\mu\nu}
\ ,
\end{align*}
where 
\begin{align*}
T^{\text{Sk}}_{\mu\nu}=&\text{Tr} \left[-c_{1}\left(R_{\mu} R_{\nu}-\frac{1}{%
2} g_{\mu \nu} R_{\alpha} R^{\alpha}\right)-\frac{c_{2}}{4}%
\left(F_{\mu}{}^{\alpha} F_{\nu \alpha}-\frac{1}{4} g_{\mu\nu} F_{\alpha
\beta} F^{\alpha \beta}\right)\right]\ , \\
T^{\rho}_{\mu\nu}=&\text{Tr}\left[-4 a_{1}\left(S_{\mu}{}^{\alpha} S_{\nu
\alpha}-\frac{g_{\mu\nu}}{4}S_{\alpha\beta} S^{\alpha\beta} \right) -2
a_{2}\left(A_{\mu} A_{\nu}-\frac{1}{2} g_{\mu \nu} A_{\alpha}
A^{\alpha}\right)\right.  \notag \\
&\left. -2 c_{3}\left(S_{\mu}{}^{\alpha} G_{\nu \alpha}+S_{\nu}{}^{\sigma}
G_{\mu \sigma}-\frac{g_{\mu \nu}}{2} S_{\alpha \beta} G^{\alpha
\beta}\right)-4 c_{4}\left(G_{\mu}{}^{\sigma} G_{\nu \sigma}-\frac{g_{\mu\nu}%
}{4} G_{\alpha \beta} G^{\alpha \beta}\right)\right] \ , \\
T^{\text{int}}_{\mu\nu}=&\text{Tr}\left[ -2\left(L_{\mu}{}^{\sigma} M_{\nu
\sigma}+L_{\nu}{}^{\sigma} M_{\mu \sigma}-\frac{g_{\mu \nu}}{2} L_{\alpha
\beta} M^{\alpha \beta}\right)+2 c_{6}\left(F_{\mu}{}^{\sigma} S_{\nu
\sigma}+ F_{\nu}{}^{\sigma} S_{\mu \sigma}-\frac{g_{\mu \nu}}{2} F_{\alpha
\beta} S^{\alpha \beta}\right) \right.  \notag \\
&\left. +2 c_{7}\left(F_{\mu}{}^{\sigma} G_{\nu \sigma}+ F_{\nu}{}^{\sigma}
G_{\mu \sigma}-\frac{g_{\mu\nu}}{2} F_{\alpha \beta} G^{\alpha \beta}\right) %
\right]\ ,
\end{align*}
with 
\begin{equation*}
M_{\mu\nu}= c_5 L_{\mu\nu}+a_{8} F_{\mu\nu}+a_9 S_{\mu\nu}+a_{10} G_{\mu\nu}
\ .
\end{equation*}
For the meron-like ansatz 
\begin{equation*}
A_{\mu }=\lambda (x)R_{\mu }\ ,
\end{equation*}%
the tensors $S_{\mu \nu }$, $G_{\mu \nu }$, $L_{\mu \nu }$ are reduced to 
\begin{equation*}
S_{\mu \nu }=\lambda F_{\nu \mu }+\nabla _{\mu }\lambda R_{\nu }-\nabla
_{\nu }\lambda R_{\mu }\ ,\quad G_{\mu \nu }=\lambda ^{2}F_{\mu \nu }\
,\quad L_{\mu \nu }=2\lambda F_{\mu \nu }\ .
\end{equation*}%
Meanwhile, $O_{1}$ and $E_{i}$ become 
\begin{align*}
O_{1}^{\mu \nu }& =P_{1}(\lambda )F^{\mu \nu }+4a_{1}(\nabla ^{\mu }\lambda
R^{\nu }-\nabla ^{\nu }\lambda R^{\mu })\ , \\
E_{1}^{\mu }& =2\left[ \lambda \frac{O_{3}^{\mu \nu }}{2}+\frac{O_{4}^{\mu
\nu }}{2},R_{\nu }\right] \ , \\
& =-(P_{2}(\lambda )-c_{2}/4)[R_{\nu },F^{\mu \nu }]-2(a_{9}\lambda
-c_{6})\nabla _{\nu }\lambda F^{\mu \nu } \\
E_{2}^{\mu }& =2\left[ \lambda \frac{O_{2}^{\mu \nu }}{2}+\frac{O_{3}^{\mu
\nu }}{2},R_{\nu }\right] \\
& =-P_{3}(\lambda )[R_{\nu },F^{\mu \nu }]-2(c_{3}\lambda +a_{9})\nabla
_{\nu }\lambda F^{\mu \nu }\ ,
\end{align*}%
where 
\begin{align*}
P_{1}(\lambda )& =2c_{3}\lambda ^{2}+4(a_{9}-a_{1})\lambda -2c_{6}\ , \\
P_{2}(\lambda )& =2a_{10}\lambda ^{3}+2(4c_{5}-a_{9}-c_{7})\lambda
^{2}+2(c_{6}+a_{8}+2a_{10})\lambda +\frac{c_{2}}{4}\ , \\
P_{3}(\lambda )& =4c_{4}\lambda ^{3}+2(3a_{10}-c_{3})\lambda
^{2}+2(a_{9}-c_{7}+2c_{5})\lambda +2a_{8}\ .
\end{align*}%
Therefore, varying the action of the Skyrme $\rho$-mesons system w.r.t the
fields $A_{\mu }$ and $U$, the fields equations turns out to be 
\begin{align*}
P_{1}(\lambda )\nabla _{\nu }F^{\mu \nu }+2a_{2}\lambda R^{\mu
}+P_{3}(\lambda )[R_{\nu },F^{\mu \nu }]+(P_{1}^{\prime }+2c_{3}\lambda
+2a_{9})\nabla _{\nu }\lambda F^{\mu \nu } & \\
+4a_{1}(\nabla _{\nu }\nabla ^{\mu }\lambda R^{\nu }+\nabla ^{\mu }\lambda
\nabla _{\nu }R^{\nu }-\Box \lambda R^{\mu }-\nabla ^{\nu }\lambda \nabla
_{\nu }R^{\mu })& =0\ , \\
c_{1}\nabla _{\mu }R^{\mu }+P_{2}(\lambda )\nabla _{\mu }[R_{\nu },F^{\mu
\nu }]-\frac{m_{\pi }c_{1}^{2}}{2c_{2}}(U+U^{-1})& \\
+(P_{2}^{\prime }-2(a_{9}\lambda -c_{6}))\nabla _{\mu }\lambda \lbrack
R_{\nu },F^{\mu \nu }]+2(a_{9}\lambda -c_{6})(\nabla _{\mu }\nabla _{\nu
}\lambda F^{\mu \nu }+\nabla _{\nu }\lambda \nabla _{\mu }F^{\mu \nu })& =0\
.
\end{align*}

\section*{Appendix B: Configurations with non-constant $\bar{\protect\lambda}
$}

\label{App2}

Here we will show a possible choice of the coupling constants of the $\rho$%
-mesons action which allows to constuct analytic solutions with a
non-constant profile $\lambda(r)$ for the $\rho$-mesons.

One can check directly that when the coupling constants are 
\begin{equation*}
c_{3}=c_{6}=c_{4}=a_{8}=a_{10}=0 \ ,\quad a_{1}=a_{9}=4c_{5}-c_{7}\ ,
\end{equation*}%
the dynamical variables $\alpha(r)$ and $\lambda(r)$ satisfy a system of
three coupled ODEs, namely 
\begin{align*}
2(c_1L^2+c_2q^2\sin^2(\alpha))\alpha^{\prime \prime
}+4(c_7-4c_5)q^2\sin^2(\alpha)(\bar{\lambda}^2)^{\prime }\alpha^{\prime }& \\
+c_2q^2\sin(2\alpha)\alpha^{\prime 2
}-2L^2(2L^2M_\pi^2+c_1q^2\cos(\alpha))\sin(\alpha) &= 0 \ , \\
\bar{\lambda}^{\prime \prime }+\frac{a_2L^2(q^2-\csc^2(\alpha)\alpha^{\prime
2})}{2q^2(c_7-4c_5)} &= 0 \ , \\
\alpha^{\prime }+\frac{q^2(c_7-4c_5)\sin(2\alpha)\bar{\lambda}^{\prime }}{%
a_2L^2\bar{\lambda}} &= 0 \ .
\end{align*}%
From the last equation we find an analytic expression for $\bar{\lambda}$ in
terms of $\alpha $, this is 
\begin{equation*}
\bar{\lambda} (r)=l_{0}\tan (\alpha )^{\frac{t_{0}}{2}} \ ,
\end{equation*}%
where $l_{0}$ is an integration constant and $%
t_{0}=a_{2}L^{2}/(4c_{5}-c_{7})q^{2}$. Once one replaces this expression for 
$\bar{\lambda}$ in the above equations two ODEs for $\alpha $ arise. The
compatibility of these equations demands that 
\begin{equation*}
c_{2}=M_{\pi }=0 \ , \qquad t_{0}=-2 \ ,
\end{equation*}%
which in turn implies 
\begin{equation*}
L=q\sqrt{\frac{2(c_{7}-4c_{5})}{a_{2}}} \ , \qquad l_{0}=\sqrt{c_{1}}/\sqrt{%
2a_{2}} \ .
\end{equation*}%
Under this conditions, the resulting equation for $\alpha $ turns out to be 
\begin{equation*}
\alpha ^{\prime \prime }-\alpha ^{\prime 2}\cot (\alpha )-q^{2}\sin (\alpha
)\cos (\alpha )=0 \ ,
\end{equation*}%
while $\bar{\lambda}$ is 
\begin{equation}
\bar{\lambda}(r) = \sqrt{c_1/2a_2}\cot{\alpha} \ .
\end{equation}
Finally, using the change of variables 
\begin{equation*}
\alpha (r)=\arccos(\tanh (H(r)))\ ,
\end{equation*}%
the equation for $\alpha$ can be leads to 
\begin{equation*}
H''+q^2 \tanh(H) \ = \ 0 \ .
\end{equation*}%

\end{document}